\documentclass[11pt,onecolumn,floatfix,altaffilletter,superscriptaddress,tightenlines,showpacs,showkeys,preprintnumbers,nofootinbib]{revtex4-2}
\pdfoutput=1
\usepackage[colorlinks=true,citecolor=blue,linkcolor=blue,breaklinks=true,linktoc=none]{hyperref}
\usepackage{amsmath,amssymb}
\usepackage{epsfig} 
\usepackage{graphicx}
\usepackage{url}
\usepackage{color}
\usepackage{multirow}
\usepackage{placeins}
\usepackage[dvipsnames]{xcolor}
\usepackage{braket}
\usepackage{float}
\usepackage{slashed}
\usepackage{comment}
\hypersetup{
  colorlinks=true,
  linkcolor=red,
  filecolor=magenta,   
  urlcolor=blue,
  citecolor=blue,
}

\hypersetup{colorlinks,linkcolor={blue},citecolor={teal},urlcolor={violet}}

\allowdisplaybreaks

\bibpunct{[}{]}{,}{n}{}{,}

\def\beq{\begin{equation}}
\def\eeq{\end{equation}}
\def\bea{\begin{eqnarray}}
\def\eea{\end{eqnarray}}
\makeatletter
\@addtoreset{equation}{section}

\newcommand{\INFN}{INFN - Sezione di Napoli, Complesso Universitario Monte S. Angelo, I-80126 Napoli, Italy}
\newcommand{\UNINA}{Dipartimento di Fisica "Ettore Pancini", Università degli studi di Napoli ``Federico II'', Complesso Universitario Monte S. Angelo, I-80126 Napoli, Italy}
\newcommand{\SSM}{Scuola Superiore Meridionale, Università degli studi di Napoli ``Federico II'', Largo San Marcellino 10, 80138 Napoli, Italy}

\begin{document}

\title{Tomography of flavoured leptogenesis with primordial blue gravitational waves}
  \author{Marco Chianese}
 \email{marco.chianese@unina.it}
 \affiliation{\UNINA}
  \affiliation{\INFN}
  \author{Satyabrata Datta}
   \email{amisatyabrata703@gmail.com}
   \affiliation{Department of Physics and Institute of Theoretical Physics,
Nanjing Normal University, Nanjing, 210023, China}
 \author{Rome Samanta}
  \email{samanta@na.infn.it}
  \affiliation{\SSM}
  \affiliation{\INFN}
 \author{Ninetta Saviano}
 \email{nsaviano@na.infn.it}
 \affiliation{\INFN}
\affiliation{\SSM}

\begin{abstract}
We explore a scenario where an early epoch of matter domination is driven by the mass scale $M_N$ of the right-handed neutrinos, which also characterizes the different flavour regimes of leptogenesis. Such a matter-domination epoch gives rise to peculiar spectral imprints on primordial Gravitational Waves (GWs) produced during inflation. We point out that the characteristic spectral features are detectable in multiple frequency bands with current and future GW experiments in case of Blue GWs (BGWs) described by a power-law with a positive spectral index $(n_T >0)$ and an amplitude compatible with Cosmic Microwave Background (CMB) measurements at the CMB scale. We find that the three-flavour leptogenesis regime with $M_N \lesssim 10^9~{\rm GeV}$ imprints BGWs more prominently than the two-flavour and one-flavour regimes characterized by a higher right-handed neutrino mass scale. {\color{black} In particular, a two-flavour (three-flavour) leptogenesis regime is expected to leave distinct imprints in the mHz-Hz ($\mu$Hz-mHz) band. Moreover, we translate the current  Big Bang Nucleosynthesis (BBN) and LIGO limits on the GW energy density into constraints on the flavour leptogenesis parameter space for different GW spectral indices $n_T$. We provide a rigorous statistical analysis of how the future GW detectors would be conjointly able to distinguish the flavour regimes.  Interestingly, the scenario also offers unique GW signals testable in the next LIGO run with a correlated signature in the PTA frequency band with an amplitude comparable to the one expected from supermassive black holes.}
\end{abstract}
\maketitle
\tableofcontents

\section{Introduction}
Primordial Gravitational Waves (GWs) can be a unique probe of physics operating at super high-energy scales that are otherwise unreachable, e.g., with terrestrial accelerators.  Broadly, a high-energy theory/model can be sensitive to GWs in two ways. First, the model can accommodate a source of GWs whose amplitude and spectral features relate to the model parameters. Thus, the model becomes testable with the GW searches.  On the other hand, the model can leave its imprints on GWs, regardless of how they are produced. This article discusses the latter case.  Let us assume a spectrum of primordial GWs, which might be of inflationary origin \cite{inf1,inf2}, and we aim at probing a post-inflation high-energy model with those GWs . In a given scenario where the underlying dynamics of GWs production are known, we expect specific spectral features in the GWs today. Therefore, testing an independent model  (which is not related to the origin of GWs) would require finding detectable imprints in those GWs caused exclusively by the model under consideration. This method proposed as a tomographic search of BSM theories with GWs \cite{Datta:2022tab,Datta:2023vbs} is similar to X-ray tomography of an object, wherein the object is placed in front of an X-ray source, and we let the X-ray pass through the object to know the invisible internal properties of the object. Likewise, the GW tomography has two main ingredients: the source, specifically the GW spectrum, and the object, namely the high-energy model to be tested.

In this article, for  GW spectrum, we consider primordial Blue Gravitational Waves (BGWs) with $\Omega_{\rm GW}\sim f^{n_T}$ (here $\Omega_{\rm GW}$ is the GWs energy density, $f$ is the frequency and $n_T>0$ is a spectral index)\footnote{As we discuss later, in a realistic setting it should be a broken power-law; otherwise the spectrum would not be cosmologically viable.} which might come, e.g., from inflation. For the high-energy model we want to probe, we consider thermal leptogenesis, which is a process that generates the baryon asymmetry of the universe~\cite{lep1}. BGWs are appealing by several considerations. First, for $n_T>0$, the BGWs come with strong amplitude, and therefore, they are easily detectable, given the projected sensitivities of the current and planned GW detectors operating at different frequency bands. Second, the recent finding of nHz stochastic GWs by the Pulsar Timing Arrays (PTAs) shows a strong preference for a blue-tilted GW spectrum of primordial origin~\cite{ng1,ng2,ng3,ng4,ng5,EPTA:2023xxk,bn1n,bn2n,bn3n,bn4n,bn5n,bn6n}. Specifically, a power-law fit to the 15 yrs NANOGrav PTA data renders a large spectral index value; $n_T \sim 1.2-2.4$ at $68\%$ CL \cite{ng5}. Third, because of their potentially strong amplitude in the sub-nHz and $\mu$Hz regions, primordial BGWs are among the very few candidates that can be tested with pulsar parameter drifts \cite{DeRocco:2022irl,DeRocco:2023qae} and possibly with Lunar Laser Ranging (LLR) \cite{Blas:2021mqw,Blas:2021mpc}. Finally, as recently pointed out, BGWs might affect large-scale structures by sourcing density perturbations at second order~\cite{Bari:2021xvf}. Although we shall not provide any model that might produce such BGWs, let us mention that the simplest single-field slow-roll inflation models produce a nearly scale-invariant spectrum with very small amplitude. Nonetheless, plenty of scenarios \cite{bgw1,bgw2,bgw3,bgw4,bgw5,bgw6,bgw7,bgw8,bgwnew,Datta:2023xpr} go beyond the simplest one and predict BGWs that are detectable at future GW experiments~\cite{ska,gnt,mrs,lisa,Baker:2019nia,bbo,decigo,Isoyama:2018rjb,et,Maggiore:2019uih,ce,ligoo5,LIGOScientific:2019hgc}.

Leptogenesis~\cite{lep1} is a two-step process: a lepton asymmetry is created in the first step, which is then processed to a baryon asymmetry via the sphaleron transition~\cite{Kuzmin:1985mm}. Among several possibilities to generate lepton asymmetry \cite{Alvarez-Gaume:1983ihn,Alexander:2004us,Cohen:1987vi}, here we consider the one within the seesaw framework of light neutrino masses. In this scenario, CP-violating and out-of-equilibrium decays of right-handed (RH) neutrinos to lepton and Higgs doublet produce lepton asymmetry. Typically, at a temperature $T\sim M_N$, where $M_N$ is the RH neutrino mass scale, lepton asymmetry gets produced. Depending on the several variants, a wide range of RH neutrino mass window, $M_N\in [\rm MeV,10^{15}\rm GeV]$ \cite{lep2,lep3,lep4,lep5,lep6,lep7,lep8,lep9,lep11,lep12}, might facilitate a successful leptogenesis. As it is obvious, $M_N\gtrsim \mathcal{O}(\rm TeV)$ scales are not reachable with collider experiments, leptogenesis operating on those scales requires alternative probes. A way to test such a high-scale scenario is provided by the flavour effects \cite{fllep1,fllep2,fllep3} which connect it indirectly to low-energy neutrino observables, specifically to the leptonic CP violation \cite{Pascoli:2006ci}. A flavoured leptogenesis scenario can have three distinct regimes (as described in the next section): one-flavour/vanilla regime (1FL)  for $M_N\gtrsim 10^{12}$ GeV, two-flavour regime (2FL) for $10^{12}$ GeV $\gtrsim M_N\gtrsim 10^{9}$ GeV, and three-flavour regime (3FL) for $M_N\lesssim 10^{9}$ GeV. Though connecting those regimes with low-energy neutrino observables is possible, it is difficult to differentiate them observationally because of plenty of free parameters in seesaw models.  One way to distinguish them is to impose further symmetries in the Lagrangian, e.g., discrete symmetries, to reduce the number of parameters \cite{fls1,fls2}. Here, we shall show that, alternatively, with the GW-tomography method, not only can such high scales be probed, but different flavour regimes can potentially leave distinct imprints on the BGW spectrum. 

\begin{figure}[t!]
    \centering
    \includegraphics[scale=1.]{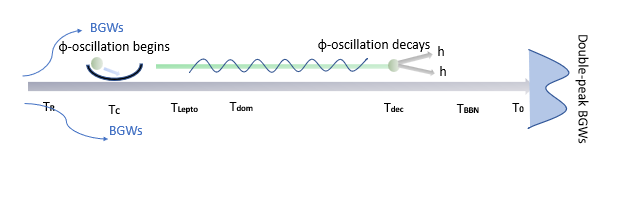}
    \caption{A schematic timeline showing the occurrence of the events considered in this work. After inflation, the universe reheats at $T=T_{\rm R}$. Phase transition happens, the scalar field oscillation begins, and RH neutrinos become massive at $T=T_c$. At the temperature $T_{\rm lepto}\simeq M_N$, the baryon asymmetry produced by the RH neutrinos freezes out. Starting from $T_{\rm dom}$ until $T_{\rm dec}$, the scalar field dominates the universe's energy density and produces entropy around $T\simeq T_{\rm dec}$. The universe becomes radiation-dominated after this. The big-bang nucleosynthesis and the current temperature are denoted by $T_{\rm BBN}$ and $T_0$, respectively.  A peaked blue-tilted GW-spectrum generated during inflation when comes across such a dynamical phase of $\Phi$, which also makes the RH neutrino massive and whose decay width is determined by the RH neutrino mass scale, becomes a double peak spectrum at present. The final spectrum carries imprints of the RH neutrino mass scale and, therefore, different regimes of leptogenesis. }
    \label{fig:history}
\end{figure}

As outlined, given an independent source of GWs, testing a model requires finding imprints on the GW spectrum caused by the model. Therefore, probing different mass scales of leptogenesis (hence different flavour regimes) requires finding characteristic GW spectral features dependent on RH masses. We show that, though nontrivial, this can be done within some parameter space in the seesaw models, which offers an RH neutrino mass-dependent matter epoch affecting the standard propagation (in radiation domination) of primordial BGWs. The idea is based on contemplating the origin of RH neutrino masses. Let us suppose that, like any other Standard Model (SM) particles, the RH neutrinos get their mass via a phase transition triggered by an SM singlet scalar field. Once the field rolls down to its vacuum and attains its vacuum expectation value, besides generating RH neutrino masses, it oscillates coherently around the vacuum. Such oscillations persist long with a null equation of state parameter (the field behaves like matter) when the scalar field lives longer. Suppose the coupling of the scalar field to the RH neutrino plays a pivotal role in determining its lifetime. In that case, the duration of the matter domination can be controlled by the coupling and, therefore, by the RH neutrino masses. This is the key aspect of our study. Irrespective of their origin, as primordial gravitational waves pass through such a matter-dominated phase, the beginning and the end of the matter-dominated phase get imprinted on the final GW spectrum.  Therefore, the final GW spectrum carries imprints of the RH neutrino mass scale.  A probable timeline in the scenario has been illustrated in Fig.~\ref{fig:history}.

Within a concrete model of $U(1)_{B-L}$ phase transition\footnote{ In the seesaw Lagrangian, $U(1)_{B-L}$ may naturally arise as a residual symmetry of many Grand Unification Theories (GUT) \cite{bml1,bml2,bml3,bml4,bml5}.} triggered by a scalar field $\Phi$ carrying $B-L$ charge, the key feature of our study is that a peaked BGW spectrum gets distorted when it passes through such an RH neutrino mass-dependent matter epoch and finally exhibits a double-peaked spectrum. The locations of the low-frequency peak ($f_{\rm low}^{\rm peak}$) and the dip ($f^{\rm dip}$) in between are strongly sensitive to RH neutrino masses. As the leptogenesis process enters from an unflavoured to flavour regimes, these two characteristic frequencies shift from a high to a low-frequency value. {\color{black} For a three-flavour regime, we can obtain the ratio $\Omega_{\rm GW}(f_{\rm low}^{\rm peak})/\Omega_{\rm GW}(f^{\rm dip})$ (the spectral distortion) much higher than the other flavour regimes. While the vanilla leptogenesis scenario hardly imprints the GW spectrum, the two-flavour and the three-flavour regimes exhibit characteristic spectral features in the GWs mostly in the mHz-Hz and $\mu$Hz-mHz bands.} The scenario can accommodate a reasonably large value of $n_T$ without contradicting any cosmological constraints and allows the overall spectrum to span a wide range of frequencies accessible to the current and planned GW detectors. Notably, in the PTA frequency band, flavoured leptogenesis regimes allow GW amplitude comparable to supermassive black hole binaries (SMBHB) with a correlated high-frequency signal testable in the next LIGO run.

{\color{black}The rest of the paper is organized as follows. In Sec.~\ref{s2}, we discuss the scalar field dynamics and all the constraints the model must comply with. In Sec.~\ref{s3}, we describe the spectral distortions induced by the model on GW signals.In Sec.~\ref{s4}, we present a detailed numerical study on how the BGW spectrum contains information on the RH neutrino mass scale and recast the current BBN and LIGO limits into bounds of the model parameter space. We also discuss our model in the context of the recent PTA results on nHz GWs and provide a rigorous statistical analysis of how the future GW detectors would be conjointly able to distinguish the flavour regimes by computing the signal-to-noise ratio (SNR). Finally, we draw our conclusions in Sec.~\ref{s5}}. 

\section{Scalar field dynamics and the scale of thermal leptogenesis} \label{s2}

The Lagrangian relevant to the discussion is given by 
\bea
-\Delta \mathcal{L}\subset f_{D i\alpha} \overline{N_{Ri} }\tilde{H}^\dagger L_{\alpha}+\frac{1}{2}f_{Ni}\overline{N_{Ri}}\Phi N_{Ri}^C+\frac{1}{4}\lambda_{H\Phi}|H|^2|\Phi|^2+V(\Phi,T)\,,\label{lag}
\eea
where $f_D$ is the neutrino Dirac Yukawa coupling, $H(L)$ is the SM Higgs (lepton doublet), $N_R$ is the RH neutrino field, $\Phi$ is a complex scalar field with vacuum expectation value $v_\Phi$, and $V(\Phi,T)$ is the finite temperature potential that determines the phase transition dynamics of $\Phi$. Here, $N_R$ and $L$ have $B-L$ charge $-1$, whereas the same charges for $\Phi$ and $H$ are $2$ and $0$, respectively. The Higgs portal coupling is in general given by $\lambda_{H\Phi}\simeq\lambda_{H\Phi}^{\rm tree}+\lambda_{H\Phi}^{\rm 1-loop}$ where $ \lambda_{H\Phi}^{\rm 1-loop}$ is generated by the first two terms in Eq.~\eqref{lag}. {\color{black} We assume $\lambda_{H\Phi}^{\rm tree} \lesssim  \lambda_{H\Phi}^{\rm 1-loop}$ on all relevant physical scales which is crucial to establish a connection between the neutrino parameters and the lifetime of $\Phi$}. The RH neutrinos become massive ($M_i=f_{Ni}v_\Phi$) when $\Phi$ obtains its vacuum expectation value $v_\Phi$. After the electroweak phase transition ($T\lesssim T_{\rm EW}$), when the Higgs settles to its minima  $H = v_h$, the first two terms generate light active neutrino mass $m_{\nu i}\sim {f_{Di}}^2v_h^2/M_i$ via the standard type-I seesaw mechanism \cite{seesaw1,seesaw2,seesaw3,seesaw4}, whereas at $T\gg T_{\rm EW}$ the heavy neutrinos decay CP asymmetrically to lepton doublets plus Higgs, and generate lepton asymmetry. At a temperature $T\sim M_i$, lepton asymmetry produced by $i$th RH neutrino freezes out; therefore, $M_i$ is typically the scale of thermal leptogenesis. For simplicity, not focusing so much on the hierarchy in $M_i$s, we consider a single scale for the rest of the discussion, i.e. $M_i\equiv M_N=f_N v_\Phi$. In the high-scale leptogenesis scenario, there are three distinct regimes of leptogenesis \cite{fllep1,fllep2,fllep3}. Generally, the RH neutrinos produce lepton doublets as a coherent superposition of flavours: $\ket{L_i} = A_{i \alpha }\ket{L_{\alpha}}$, where $\alpha = e,\mu,\tau$ and $ A_{i\alpha }$ is the corresponding amplitude determined by $f_{D i\alpha}$. For temperatures $T\gtrsim 10^{12}~{\rm GeV}$, all the charged lepton Yukawa couplings are out of equilibrium (the interaction strength is given by $\Gamma_\alpha\sim 5\times 10^{-3}~f_{L\alpha}T$, with $f_{L\alpha}$ being the charged lepton Yukawa coupling) and do not participate in the leptogenesis process. This is the one flavour or the vanilla regime of leptogenesis (1FL). In this regime, leptogenesis is not sensitive to the neutrino mixing matrix, e.g. mixing angles and low-energy CP phases. When the temperature drops down to $T\lesssim10^{12}~{\rm GeV}$, the $\tau$ flavoured charged lepton interaction comes into equilibrium and breaks the coherent $\ket{L_i}$ state into $\tau$ and a composition of $e+\mu$ flavour (two flavour regime, 2FL). Likewise, the later composition is broken once $T\lesssim 10^9~{\rm GeV}$, when the $\mu$ flavoured charged lepton interaction comes into equilibrium (three flavour regime, 3FL). In both these flavoured regimes, leptogenesis may be sensitive to the low-energy neutrino parameters. We will explore all these regimes and show that the flavoured regime of leptogenesis leaves much more distinct imprints on BGWs than the vanilla one, paving the path for a possible synergy between low-energy neutrino physics and GWs.

As the temperature drops, the scalar field transits from $\Phi=0$ towards its vacuum expectation value $\Phi=v_{\Phi}$. The finite temperature potential that restores the symmetry at higher temperatures is given by \cite{Linde:1978px,Kibble:1980mv,Quiros:1999jp,Caprini:2015zlo,Hindmarsh:2020hop}
\bea
V(\Phi,T)=D(T^2-T_0^2)\Phi^2-ET\Phi^3 + \frac{\lambda}{4}\Phi^4\,,\label{tmpdp}
\eea
with
\bea
D=\frac{3{ g^\prime}^2+4\lambda}{24}\,,~~E=\frac{3{ g^\prime}^3+{ g^\prime}\lambda+3\lambda^{3/2}}{24 \pi}\,, ~~ T_0=\frac{ \sqrt{12\lambda}v_{\Phi}}{\sqrt{3{ g^\prime}^2+4\lambda}}\,,
\eea 
where $g^\prime$ is the gauge coupling,\footnote{As mentioned in the introduction, seesaw models with $U(1)_{B-L}$ gauge symmetry is well-motivated from GUT. In that case, the scalar field carries a gauge charge. Therefore, the gauge coupling appears in the finite temperature potential.} and the vacuum expectation value $v_\Phi=\mu/\sqrt{\lambda}$ has been determined from the zero temperature potential $V(\Phi,0)=-\frac{\mu^2}{2}\Phi^2+\frac{\lambda}{4}\Phi^4$. The structure of the finite temperature potential determines how the transition proceeds. The last term in Eq.~\eqref{tmpdp} generates a potential barrier causing a secondary minimum at $\Phi\neq0$, which at $T=T_c$ becomes degenerate with the $\Phi=0$ one. At $T_0 ~(\lesssim T_c)$, the potential barrier vanishes, making the minimum at $\Phi=0$ a maximum~\cite{Quiros:1999jp}. The critical temperature $T_c$ and the field value $\Phi_c = \Phi \left(T_c\right)$ are given by~\cite{Quiros:1999jp,Megevand:2016lpr}
\bea
T_c=T_0\frac{\sqrt{\lambda D}}{\sqrt{\lambda D-E^2}}\,,~~ \Phi_c = \sqrt{\frac{4 D}{\lambda}(T_c^2-T_0^2)}\,.
\eea
A non-zero value of $E$ in Eq.~\eqref{tmpdp} generally leads to a first-order transition with a strength determined roughly by the order parameter $\Phi_c/T_c$ \cite{Quiros:1999jp}. Nonetheless, if $\Phi_c/T_c\ll 1$, the transition is extremely weakly first-order, which can be treated as a 'second-order' transition because the potential barrier disappears quickly ($T_c \simeq T_0$). In this case, the transition can be described by rolling of the field $\Phi$ from $\Phi=0$ to $\Phi = v_\Phi$. In this article, we consider such a case working with the values of $\lambda$ and  ${g^\prime}$ so that $\Phi_c/T_c\ll 1$ is fulfilled. In particular, we find that, for ${g^\prime}^3 \lesssim \lambda \lesssim 1$ and ${g^\prime} \lesssim 10^{-2}$, the order parameter is $\Phi_c/T_c\lesssim 0.08$.\footnote{In Ref.\cite{Datta:2022tab}, it was checked numerically that the transition is a second-order type for this choice of parameters.} Fixing the hierarchy between the couplings to be $\lambda \simeq {g^\prime}^3$ (limiting condition), we also have $T_c \simeq T_0 = 2\sqrt{g^\prime} v_\Phi$. {\color{black} Additionally, not to have a second period of inflation and for the field to roll, the conditions $\rho_\Phi (T_c)<\rho_R(T_c)$ and $m_\Phi=\sqrt{2\lambda} v_\Phi\gtrsim \mathcal{H}(T_c)$ must be satisfied, where $\rho_i$ is the energy density and $\mathcal{H}$ is the Hubble parameter. In terms of our model parameters, these conditions then translate to  
\bea
\rho_\Phi (T_c)<4.5\times 10^{-6}\left(\frac{g^\prime}{10^{-2}}\right)\rho_R(T_c)\,,\,\,{\rm and}\,\,\left(\frac{g^\prime}{10^{-2}}\right)^{1/2}\gtrsim 4\times 10^{-4}\left(\frac{v_\Phi}{10^{13}\,\rm GeV}\right)\,,\label{const1}
\eea
where the initial energy density of the scalar field is given by $\rho_{\Phi}\left(T_c \right) \simeq \lambda v_\Phi^4/4$.

Once the field rolls down to the true vacuum, it oscillates around $v_{\Phi}$. For a generic potential  $V(\Phi)=\alpha \Phi^\beta$, the equation of state of such a coherent oscillation can be computed as~\cite{Datta:2022tab}
\bea
\omega=(\beta-2)(\beta+2)^{-1}\,.\label{eosb}
\eea
Assuming the oscillation of the scalar field is driven by the dominant quadratic term in the potential and expanding the zero temperature potential around the true vacuum, we obtain $\alpha=\lambda v_{\Phi}^2$ and $\beta=2$. Therefore, the scalar field behaves like matter ($\omega=0$). We can also compute the angular frequency of oscillation, which is $m_\Phi=\sqrt{2\lambda} v_\Phi$. If $\Phi$ is long-lived (lifetime set by the decay width as discussed below), these oscillations persist and the universe goes through a significant period of matter domination, thus leading to distinct imprints on the BGWs as we discuss later.

Let us now constrain the RH neutrino mass scale (scale of leptogenesis) $M_N=f_N v_\Phi$ and the decay width of $\Phi$. First, the RH neutrino Yukawa coupling $f_N$ should be large because we consider high-scale leptogenesis. However, in that case, if $m_\Phi > 2 M_N$, $\Phi$ decay to RH neutrino pairs ($\Phi\rightarrow N N$) would be too quick to provide matter domination. Therefore, we restrict ourselves to the case $M_N\gtrsim m_\Phi$. Second, RH neutrinos become massive after the phase transition at $T=T_c$. Therefore, the scale of leptogenesis $T_{\rm lepto} \sim M_N$ is bounded from above as $T_{\rm lepto} \sim M_N\lesssim T_c$. For $\lambda\simeq {g^\prime}^3$, the constraint $m_\Phi\lesssim M_N\lesssim T_c$ from requiring long-lived $\Phi$ and successful leptogenesis thus becomes\footnote{In numerical computation, we shall consider a bit stronger upper limit: $f_N\lesssim g^{\prime 3/4}$, by neglecting the contribution of  the terms $\mathcal{O}(f_N^4)$ in the effective potential.} 
\bea
\sqrt{2g^\prime}g^\prime \lesssim f_N\lesssim 2\sqrt{g^\prime}\,.\label{rhmass_b}
\eea
Notice that in this specific case, the mass of $\Phi$ is suppressed by a factor $\sqrt{g^\prime}$ compared to the $M_{Z^\prime}=\sqrt{2}g^\prime v_\Phi$, thus kinematically forbidding the $\Phi$ decays into $Z^\prime$ pairs.

Two competitive decay channels of $\Phi$ are now $\Phi\rightarrow hh$ and $\Phi\rightarrow f\bar{f}V$, where $h$, $f$, and $V$ are the SM Higgs, leptons and vector bosons, respectively (more about $\Phi$ decays are discussed at the end of this section). Both the decays are radiative and therefore suppressed, making $\Phi$ long-lived. The first decay channel, $\Phi\rightarrow hh$, is mediated by the neutrino Dirac Yukawa $f_D$ and the coupling $f_N$ (see Fig.~\ref{fig:loop}). Hence, it is directly connected to light neutrino and leptogenesis parameter space, making it the main decay channel in our study. In the limit $m_\Phi \gg m_h$, the decay rate is given by \cite{Gross:2015bea,Enqvist:2016mqj,Croon:2019dfw}
\begin{equation}
\Gamma_\Phi^{hh}\simeq \frac{({\lambda_{H\Phi}^{\rm 1-loop}})^2 \, v_\Phi^2}{32\pi m_\Phi} \simeq \Gamma_0~\frac{f_N^6}{\lambda} \left(\frac{v_\Phi}{\rm 10^{13}~GeV}\right)^2\left(\frac{m_\Phi}{\rm 10^8~GeV}\right)~\ln^2\left(\frac{\Lambda}{\mu}\right)\,,\label{newgamma}
\end{equation}
where the one-loop effective coupling $\lambda_{H\Phi}^{\rm 1-loop} \sim f_D^2 f_N^2 /(2\pi^2)\ln \left({\Lambda}/{\mu}\right)$  with $f_D=\sqrt{m_\nu f_N v_\Phi/v_h^2}$ has been evaluated for $m_\nu\simeq 0.01$~eV plus $v_h=174$~GeV, and $\Gamma_0\simeq 1.3\times 10^{-2}$. The logarithmic factor accounts for the renormalization condition such that at a very high scale, e.g. at $\Lambda\equiv\Lambda_{\rm GUT}\simeq 10^{16}~{\rm GeV}$,\footnote{Note that $\Lambda$ is a free scale. For phenomenological purposes, it is therefore sufficient to consider non-vanishing effective coupling at low energies to obtain matter domination and consequently detectable spectral distortion in the GWs. In our scenario, $\Lambda\simeq T_{\rm lepto}\gtrsim m_\Phi$ would not change the result significantly because the dependence is logarithmic. We can therefore also safely neglect any Higgs portal physics at temperatures $T\gtrsim T_{\rm lepto}\gtrsim m_\Phi$ } the 1-loop portal coupling vanishes, and $\mu$ is a relevant physical scale. Considering $\mathcal{H}\simeq \Gamma_\Phi^{hh}$ we can now compute $T_{\rm dec}$ as 
\bea
T_{\rm dec}=\widetilde{T}_{\rm dec} \ln\left(\frac{\Lambda}{\mu}\right)\,,\label{eq:tdec}
\eea
where 
\bea
\widetilde{T}_{\rm dec}=\left(\frac{90}{\pi^2 g_*}\right)^{1/4} \left[\tilde{M}_{\rm Pl}~\Gamma_0~\frac{f_N^6}{\lambda}\left(\frac{v_\Phi}{\rm 10^{13}~GeV}\right)^2\left(\frac{m_\Phi}{\rm 10^8~GeV}\right)\right]^{1/2}\,,
\eea
with $\tilde{M}_{\rm Pl}=2.4\times 10^{18}~{\rm  GeV}$ being the reduced Planck constant, and $g_* \simeq 106.75$ the effective degrees of freedom that contribute to the radiation. For the physical scale $\mu\equiv T_{\rm dec}$, the solution to Eq.~\eqref{eq:tdec} is given by
\bea
T_{\rm dec} = \tilde{T}_{\rm dec}~\mathcal{W}\left(\frac{\Lambda}{\widetilde{T}_{\rm dec}}\right)\,, \label{newdec}
\eea
where $\mathcal{W} ({\Lambda}/{\widetilde{T}_{\rm dec}})$ is a Lambert function that modulates the $T_{\rm dec}$ owing to the imposed renormalisation condition. The Lambert function and the coupling $f_N$ have opposite effects on $T_{\rm dec}$. The former tends to increase $T_{\rm dec}$ because $\mathcal{W} ({\Lambda}/{\widetilde{T}_{\rm dec}})\gg 1$ for $\Lambda\gg \widetilde{T}_{\rm dec}$ (which is the case in our scenario), whereas a decrease in the latter results in a decrease in $T_{\rm dec}$. However, since the quantity $\widetilde{T}_{\rm dec}$ goes as $f_N^3$, the latter effect is much stronger. In addition to the conditions described in Eq.s~\eqref{const1} and~\eqref{rhmass_b}, a viable parameter space of the framework must also comply with the following constraints: first, the scalar field should decay before the BBN, i.e., $T_{\rm dec}\gtrsim T_{\rm BBN}\sim 10 ~ {\rm MeV}$; second, the channel $\Phi\rightarrow hh$ has to dominate over $\Phi\rightarrow f\bar{f}V$ (we need the matter-dominated epoch driven by $\Phi$ to be determined by the neutrino parameters), i.e, $\Gamma_\Phi^{hh}\gtrsim \Gamma_\Phi^{f\bar{f}V}$, where $\Gamma_\Phi^{f\bar{f}V}$ is given by~\cite{Blasi:2020wpy}
\begin{figure}[t!]
    \centering
    \includegraphics[scale=1.1]{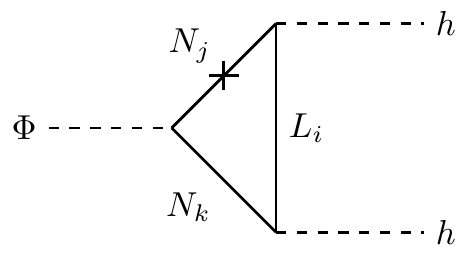}
    \caption{Diagram generating leading-order radiative $\Phi H H$ coupling through RH neutrino and Lepton mediation~\cite{Blasi:2020wpy,Gross:2015bea,Enqvist:2016mqj,Croon:2019dfw}.}
    \label{fig:loop}
\end{figure}
\bea
\Gamma_\Phi^{f\bar{f}V}\simeq \lambda{g^\prime}^4\left(\frac{m_\Phi}{\rm 10^8~GeV}\right)\,.
\eea
Given the above constraints, we now extract the viable parameter space of our model and compute the quantities $T_{\rm dec}$, $T_{\rm dom}$ and the entropy production factor $\kappa$ which are most relevant to describe the spectral features in the BGWs discussed in the next section. In the computation of GW spectra, we use analytical expressions for these quantities. While $T_{\rm dec}$ is defined in Eq.~\eqref{newdec}, the other two are given by
\bea
T_{\rm dom}=\frac{\rho_\Phi(T_c)}{\rho_R(T_c)}T_c\,,\,\,{\rm and}\,\,\kappa=\frac{T_{\rm dom}}{T_{\rm dec}}\Theta\left[T_{\rm dom}-T_{\rm dec}\right] \,,\label{newentr}
\eea
where the $\Theta$ function represents that the expression for $\kappa$ is valid for $T_{\rm dom}>T_{\rm dec}$ only.

To crosscheck the accuracy of the analytical results and to track the evolution of the energy densities and the entropy production as a function of the temperature (see e.g. the right panel of Fig.~\ref{fig:model}), it is useful to solve the following Friedmann equations for the energy densities: 
\bea
\frac{{\rm d}\rho_R}{{\rm d}t}+4\mathcal{H}\rho_R=\Gamma_\Phi^{hh}\rho_{\Phi}\,,~~
\frac{{\rm d}\rho_{\Phi}}{{\rm d}t}+3\mathcal{H}\rho_{\Phi}=-\Gamma_\Phi^{hh}\rho_{\Phi}\,,~~
\frac{{\rm d}s}{{\rm d}t}+3\mathcal{H}s=\Gamma_\Phi^{hh}\frac{\rho_{\Phi}}{T}\,,\label{be3}
\eea
upon recasting them as
 \bea
\frac{{\rm d}\rho_{R}}{{\rm d}z}+\frac{4}{z}\rho_R=0\,, ~~
\frac{{\rm d}\rho_{\Phi}}{{\rm d}z}+\frac{3}{z}\frac{\mathcal{H}}{\tilde{\mathcal{H}}}\rho_{\Phi}+\Gamma_\Phi^{hh}\frac{1}{z\tilde{\mathcal{H}}}\rho_{\Phi}=0\,,\label{den2}
\eea
where $z=T_c/T$, $s$ is the entropy density of the thermal bath, and from the third of Eq.~\eqref{be3}, the temperature-time relation has been derived as
\bea
\frac{1}{T}\frac{{\rm d}T}{{\rm d}t}=-\left(\mathcal{H}+\frac{1}{3g_{*s}(T)}\frac{{\rm d}g_{*s}(T)}{{\rm d}t}-\Gamma_\Phi^{hh}\frac{\rho_{\Phi}}{4\rho_{R}}\right)=-\tilde{\mathcal{H}}\,.\label{temvar}
\eea
On the other hand, the amount of entropy production from the $\Phi$ decays can be computed by solving 
\bea
\frac{{\rm d}a}{{\rm d}z}=\left(1+\Gamma_\Phi^{hh}\frac{\rho_{\Phi}}{4\rho_{R}\tilde{\mathcal{H}}}\right)\frac{a}{z}\,,
\eea
with $a$ being the scale factor, and then computing the ratio of $\tilde{S}\sim a^3/z^3$, i.e.,  $\kappa=\tilde{S}_{\rm after}/\tilde{S}_{\rm before}$.
\begin{table}[t!]
    \centering
    \begin{tabular}{c|c|c|c|c|c|c}
         Benchmark cases& $~~M_N~[{\rm GeV}]~~$ & $f_N$ & $~~~v_\Phi~[{\rm GeV}]~~~$ & $~~T_{\rm dom}~[{\rm GeV}]~~$& $~~T_{\rm dec}~[{\rm GeV}]~~$& $~~~~~~\kappa~~~~~~$\\ \hline
         BP1 (1FL) & $1.4\times 10^{12}$ & $~~2.0\times 10^{-3}~~$ & $  7.0\times 10^{14}$ & $ 6.4 \times 10^8$ & $ 9.8 \times 10^7$ & $ 6.5$\\
         BP2 (2FL) & $1.6\times 10^{10}$ & $~~1.4\times 10^{-3}~~$ & $ 1.1 \times 10^{13}$ & $  1.0 \times 10^7$& $ 1.0 \times 10^5$& $ 1.0 \times 10^2$\\
         BP3a (3FL) & $1.3\times 10^{8}$ & $~~3.4\times 10^{-3}~~$ & $ 3.9 \times 10^{10}$ & $ 3.4 \times 10^4$ & $ 3.4 \times 10^2$ & $ 1.0 \times 10^2$\\
         BP3b (3FL) & $1.3\times 10^{8}$ & $~~1.4\times10^{-3}~~$ & $ 9.5 \times 10^{10}$ & $ 8.6\times 10^4$& $ 9.7\times 10^1$& $ 8.9\times 10^2$\\
    \end{tabular}
    \caption{The four benchmark points considered in this paper.  In each case, leptogenesis occurs in the standard radiation domination because $T_{\rm lepto}\sim M_N>T_{\rm dom}$, where $T_{\rm dom}$ is the temperature at which the scalar field starts to dominate. }
    \label{tab:BP}
\end{table}

In the most general setting, our model has four independent parameters which are the $\Phi$ vacuum expectation value $v_\Phi$ and the three couplings $f_N$, $\lambda$ and $g^\prime$ (or equivalently the three masses $M_N$, $m_\Phi$ and $M_{Z^\prime}$). However, for the sake of definiteness, we hereafter assume $\lambda \simeq g^{\prime 3}$ which is the limiting condition to achieve a second-order phase transition. Concerning the gauge coupling, although we do not specify any GUT scheme, large values of $g^\prime$ are generally desirable for realistic GUT scenarios. In the main analysis, we discuss the results for $g^\prime = 10^{-2}$ that captures the qualitative features of the parameter space. In the Appendix~\ref{app:1}, we investigate the impact of varying $g^\prime$ within an order of magnitude and show that for an appreciable range of $g^\prime$, the qualitative results of our analysis remain the same.
\begin{figure}[t!]
\includegraphics[width=0.59\textwidth]{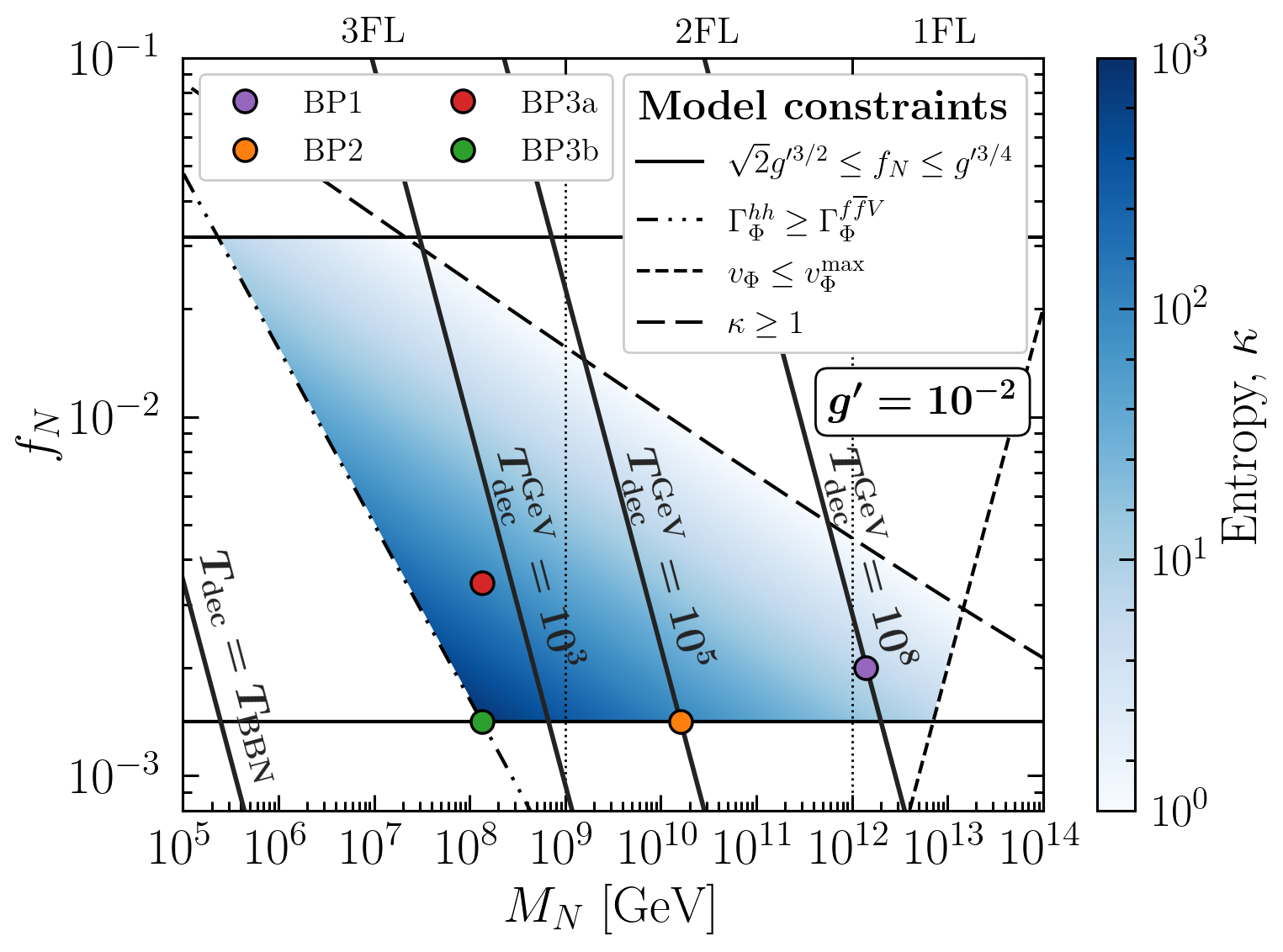}
\includegraphics[width=0.39\textwidth]{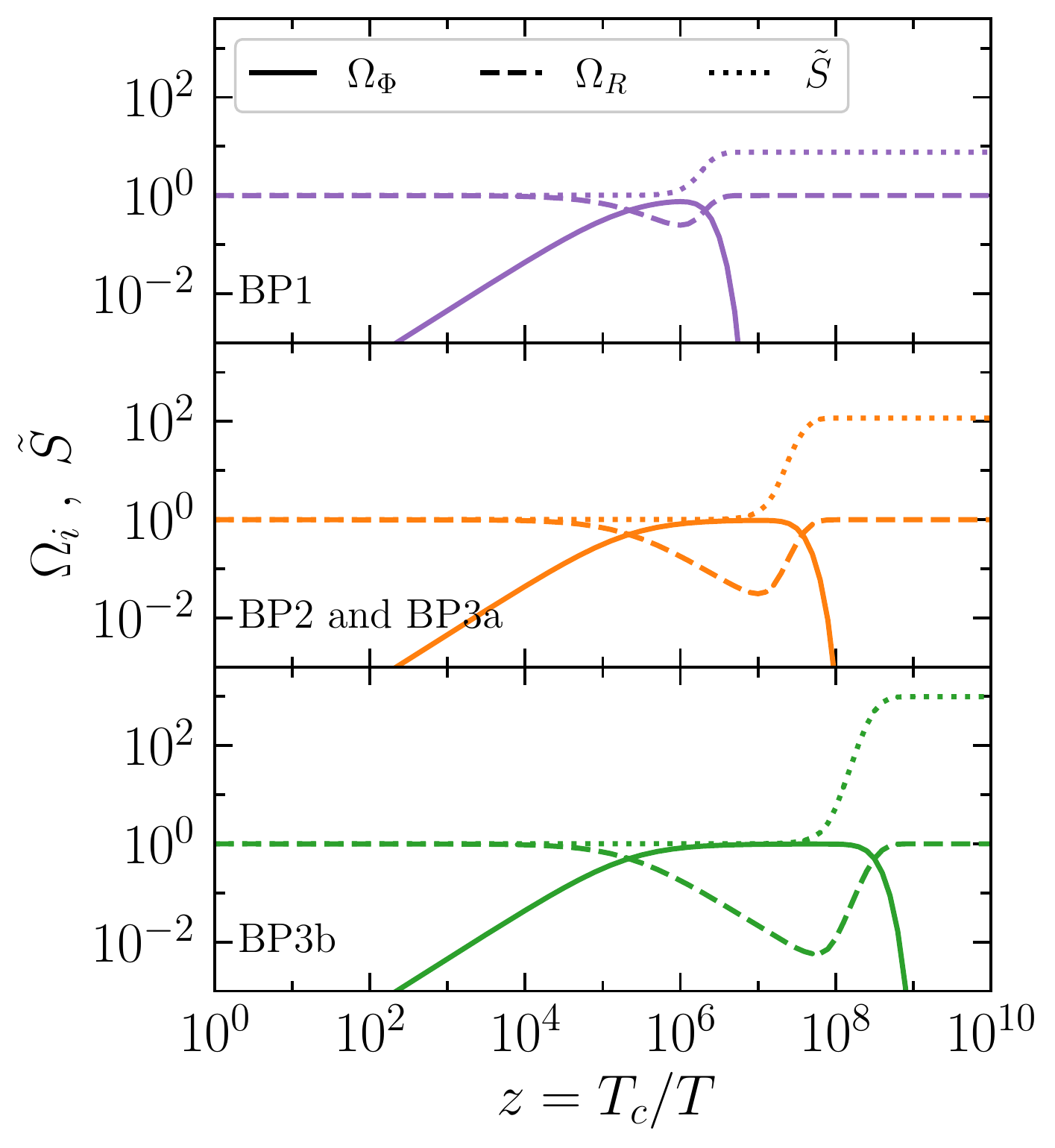}
\caption{Left: The entropy produced ($\kappa$) by the $\Phi$ decays on the $M_N$-$f_N$ plane for $g^\prime = 10^{-2}$. The white regions are excluded by the different constraints represented by the thin black lines (for more details see Sec.~\ref{s2}). The oblique thick black lines are contours for different values of $T_{\rm dec}$ (in units of GeV), while the dotted vertical lines highlight the transition among the different flavour regimes of leptogenesis. The four points correspond to the four benchmark cases reported in Tab.~\ref{tab:BP}.
Right: the evolution of the normalized energy density $\Omega_i = \rho_i/\rho_{\rm tot}$ for $\Phi$ scalar (solid lines) and radiation (dashed lines), and the total entropy (dotted lines) as a function of the auxiliary variable $z$. The three panels are for the benchmark points shown in the left plot. The benchmark points BP2 and BP3a share the same plots since the ratios among the temperatures $T_c$, $T_{\rm dom}$ and $T_{\rm dec}$ are the same (see also Tab.~\ref{tab:BP}). Because only the ratio of the entropy after and before the decay is important, the initial value of $\tilde{S}$ has been chosen as $\tilde{S}(z=1)=1$. The parameter $\kappa$ in Eq.~\eqref{newentr} is therefore $\kappa=\tilde{S}_{\rm after} = \tilde{S}\left(z\rightarrow \infty\right)$.} \label{fig:model} 
\end{figure}
 
In Fig.~\ref{fig:model} we show the produced entropy on the allowed model parameter space (left plot) and the evolution of the normalised energy densities $\Omega_i=\rho_i/\rho_{\rm tot}$ and the entropy\footnote{ We have also checked that the analytical expression for the entropy production factor $\kappa$ in Eq.~\eqref{newentr} matches the numerical result (Fig.~\ref{fig:model}, right) with good accuracy.} (right plot) for the four sets of benchmark parameters (see Tab.~\ref{tab:BP}). The benchmark points are chosen such that each falls into a specific regime of leptogenesis: BP (1, 2, 3) are in 1FL, 2FL, and 3FL regimes, respectively. Note that BP3a and BP2 fall into the category of equal entropy production. Despite this, as shown in the next section, they exhibit different spectral features because they correspond to different $T_{\rm dec\,(dom)}$. On the other hand, BP3b corresponds to the largest entropy production in this model which is exclusive only to the 3FL and produces a very distinct GW signal.

In the left panel of Fig.~\ref{fig:model}, the Yukawa coupling $f_N$ is bounded from above and below. The lower bound comes from Eq.~\eqref{rhmass_b} while the upper bound corresponds to $f_N^4\lesssim {g^\prime}^3$ because we neglected the terms $\mathcal{O}(f_N^4)$ in the effective potential. This is a bit stronger upper bound than the one shown in Eq.~\eqref{rhmass_b}. The white triangular region on the top is excluded because $T_{\rm dom}<T_{\rm dec}$ ($\kappa < 1$), which is unphysical. The region below the dot-dashed line corresponds to $\Gamma_\Phi^{f\bar{f}V}>\Gamma_\Phi^{hh}$ which is excluded. The region on the right of the short-dashed line is instead excluded since $v_\Phi$ is larger than $v^{\rm max}_\Phi \simeq 5 \times 10^{15}~{\rm GeV} \lesssim \Lambda_{\rm GUT}$. The parameter space presented in Fig.~\ref{fig:model} is also consistent with the other two constraints reported in Eq.~\eqref{const1}. The oblique black thick lines correspond to different decay temperatures. Notice that the different flavour regimes of leptogenesis correspond to very distinct ranges for $T_{\rm dec}$ which are: $T_{\rm dec} \lesssim 10^{3}~{\rm GeV}$ for 3FL, $10^{5}~{\rm GeV}\lesssim T_{\rm dec} \lesssim 10^{8}~{\rm GeV}$ for 2FL, and $T_{\rm dec} \gtrsim 10^{8}~{\rm GeV}$ for 1FL. As we will show later, this distinction will be more pronounced when looking at the spectral distortion in the GWs.

We conclude this section with the following remarks. First, as we shall see in the next section, larger entropy production corresponds to more prominent spectral distortion in the GWs. Therefore, the region around $M_N\simeq m_\Phi$ is the best parameter space of this model. Second, this paper does not present any detailed computation related to the baryon asymmetry. We however note that the baryon asymmetry produced by the RH neutrinos will be diluted due to the entropy injection given in Eq.~\eqref{newentr}. Therefore, unlike the conventional choice of a strong hierarchy in the RH neutrino masses, a more suitable option would be to make RH neutrinos quasi-degenerate. This not only evades the issue of entropy dilution by enhancing the baryon asymmetry via the resonant-leptogenesis mechanism~\cite{lep3}, but also justifies our choice of a single RH neutrino mass scale $M_i\equiv M_N$ throughout. We can get an order of magnitude estimate on the level of quasi-degeneracy in the RH neutrinos. The produced baryon-to-photon ratio in the resonance regime can be written as~\cite{lep3} 
\bea
\eta_B =10^{-2}\varepsilon_i\zeta_i \kappa^{-1}\,\,\, {\rm with } \,\,\,
\varepsilon_i\simeq\frac{3M_i m_\nu}{8\pi v_h^2\delta}\,,\label{bas}
\eea
where $\varepsilon_i$ is the CP asymmetry parameter,  $\zeta_i\simeq 10^{-1}$ is the efficiency of lepton asymmetry production, $\delta=(M_j-M_i)/{M_i}$, and $\kappa$ is the injected entropy by $\Phi$ decay as defined in Eq.~\eqref{newentr}. We can obtain an expression for $\delta$ in two cases. The standard case without entropy production corresponds to $\kappa=1$. Given the observed value $\eta_B\simeq 6.3\times 10^{-10}$ \cite{Planck:2015fie}, using Eq.~\eqref{bas}, we obtain 
\bea
\delta_S \simeq  6.25\times 10^{-1} \left(\frac{M_N}{10^{9}\rm GeV}\right)\,,\,\,\delta_E\simeq \delta_S \kappa^{-1},\\
\eea
where the subscript $S(E)$ represents the mass splitting in the standard (with entropy production) case. Therefore, e.g., for BP3b, we get $\delta_E$ to be three orders of magnitude (see the value of $\kappa$ in Tab.~\ref{tab:BP}) less than the standard case. Second, we assume $\lambda_{H\Phi}^{\rm 1-loop}\gtrsim \lambda_{H\Phi}^{\rm tree}$ for the thermal history to be determined by the neutrino parameters. Therefore, for our scenario to work, we must have 
\bea
\lambda_{H\Phi}^{\rm tree}\lesssim \left[\Gamma_\Phi^{hh} \frac{32 \pi m_\Phi}{v_\Phi^2}\right]^{1/2}.
\eea
Using Eq.~\eqref{newgamma} and the chosen values of $f_N$, e.g., for BP2 (BP3a), we have $\lambda_{H\Phi}^{\rm tree} \lesssim 10^{-11(12)}$. This shows that, in either case, the $B-L$ scalar must be feebly coupled with the SM Higgs at the tree level.  Finally, let us also remark that, while virtual $Z^\prime$ mediated 1-loop decays of $\Phi$ to two fermion final states are chirality suppressed \cite{Blasi:2020wpy,Han:2017yhy} compared to  $\Phi\rightarrow f\bar{f}V$, $N_R$ and $Z^\prime$ mediated 1-loop decays with $N_R$ and $Z^\prime$s at the final states are not allowed because of the constraint in Eq.~\eqref{rhmass_b}. Another potentially interesting decay channel $\Phi\rightarrow ZZ$  might pop up due to kinetic mixing between $U(1)_Y$ and $B-L$ gauge bosons. In the $U(1)_{B-L}$ model, the minimum size of the kinetic mixing obtained at 1-loop level is given by $\epsilon \sim g g^\prime/16\pi^2$ (up to some logarithmic factor imposed by the renormalization condition)~\cite{km1,km2,km3}, where $g$  and $g^\prime$ are the couplings of two $U(1)$'s. Therefore for large  $U(1)_{B-L}$ coupling, $\epsilon$ could be large--in fact, this is the reason that kinetic mixing imposes significant restriction on the parameter space of low mass $Z^\prime$ searches \cite{km4}. In our case, however, $Z^\prime$ is extremely heavy and mixes with the SM gauge boson eigenstate with a strength $\alpha\sim {\rm sin \,\theta_W}\,\epsilon\,({m_Z}/{M_{Z^\prime}})^2$, where $\theta_W$ is the weak mixing angle. Because of this suppressed mixing,  even for $g^\prime \sim  10^{-2}$, the induced $\Phi\Phi Z Z$ coupling is much weaker than that of $\Phi\Phi h h$ used in our analysis. But this effect could be considerable in a top-down model (e.g. if the model predicts large $g^\prime$ and small vacuum expectation value of the $B-L$ scalar).

\section{Imprints of flavour regimes of leptogenesis on BGWs} \label{s3}

In what follows, we briefly discuss the GW production during inflation and their propagation through multiple cosmological epochs until today. The perturbed FLRW line element that describes GWs is given by
\bea
{\rm d}s^2=a(\tau)\left[-{\rm d}\tau^2+(\delta_{ij}+h_{ij}){\rm d}x^{\rm i}dx^j)\right]\,,
\eea
where $\tau$ is the conformal time, and $a(\tau)$ is the scale factor. The transverse and traceless ($\partial_ih^{ij}=0$, $\delta^{ij}h_{ij}=0$) part of  $h_{ij}$ represents the GWs. Because the GWs are weak, $|h_{ij}|\ll1$, the following linearized evolution equation 
\bea
\partial_\mu(\sqrt{-g}\partial^\mu h_{ij})=16\pi a^2(\tau) \mathcal{\pi}_{ij}\label{lineq}
\eea
is sufficient to study the propagation of the GWs. The quantity $\pi_{ij}$ represents an anisotropy stress tensor that couples to $h_{ij}$ as an external source, which within a realistic cosmic setting only affects the GWs at large scales (than those of PTAs), e.g., due to the free streaming of light neutrinos~\cite{Weinberg:2003ur,Zhao:2009we}. The Fourier space decomposition of $h_{ij}$ reads  
\bea
h_{ij}(\tau, \vec{x})=\sum_\lambda\int \frac{{\rm d}^3\vec{k}}{(2\pi)^{3/2}} e^{i\vec{k}.\vec{x}}\epsilon_{ij}^\lambda(\vec{k})h_{\vec{k}}^\lambda(\tau)\,,\label{fug}
\eea
where the index $\lambda=``+/-"$ represents two polarisation states. The polarisation tensor $\epsilon_{ij}^\lambda(\vec{k})$ besides being transverse and traceless, complies with the following conditions: $\epsilon^{(\lambda)ij}(\vec{k})\epsilon_{ij}^{(\lambda^\prime)}(\vec{k})=2\delta_{\lambda\lambda^\prime}$ and $\epsilon^{(\lambda)}_{ij}(-\vec{k})=\epsilon^{(\lambda)}_{ij}(\vec{k})$. Assuming isotropy and similar evolution of each polarisation state, we replace $h_{\vec{k}}^\lambda(\tau)$ with $h_{k}(\tau)$, where $k=|\vec{k}|=2\pi f$ with $f$ being the present frequency at $a_0=1$. Neglecting subdominant contribution from $\pi_{ij}$, the GW propagation equation in the  Fourier space reads
\bea
\ddot{h}_k+2\frac{\dot{a}}{a}\dot{h}_k+k^2h_k=0\,, \label{prpeq}
\eea
where the dot represents a conformal time derivative. Using Eq.~\eqref{fug} and Eq.~\eqref{prpeq}, the energy density of the GWs can be computed as~\cite{WMAP:2006rnx}
\bea
\rho_{\rm GW}=\frac{1}{32\pi G}\int\frac{{\rm d}k}{k}\left(\frac{k}{a}\right)^2T_T^2(\tau, k)P_T(k)\,,\label{gw1}
\eea
where $T_T^2(\tau, k)=|h_k(\tau)|^2/|h_k(\tau_i)|^2$ is a transfer function  with $\tau_i$ as the initial conformal time. The quantity $P_T(k)=k^3|h_k(\tau_i)|^2 / \pi^2$ characterizes the primordial power spectrum and connects to the inflation models with specific forms. Here we opt for a  power-law parametrization of $P_T$:
\bea
P_T(k)=r A_s(k_*)\left(\frac{k}{k_*}\right)^{n_T}\,,\label{ptp}
\eea
where $r\lesssim 0.06$~\cite{BICEP2:2018kqh} is the tensor-to-scalar-ratio, and $A_s \simeq 2\times 10^{-9}$ is the scalar perturbation amplitude at the pivot scale $k_*=0.01~{\rm  Mpc^{-1}}$. In the present analysis, we take $r=0.06$ and consider the tensor spectral index $n_T$ constant plus blue-tilted ($n_T>0$). However, a scale dependence might arise owing to the higher order corrections~\cite{Kuroyanagi:2011iw,Calcagni:2020tvw} depending on the inflation model, which we do not discuss here.\footnote{We point out that the single field slow-roll inflation models correspond to the consistency relation $n_T=-r/8$~\cite{Liddle:1993fq}, i.e., the spectral index is slightly red-tilted~($n_T\lesssim 0$).} The normalized GW energy density  pertinent to detection purposes is expressed as 
\bea
\Omega_{\rm GW}(k)=\frac{k}{\rho_c}\frac{{\rm d}\rho_{\rm GW}}{{\rm d}k}\,,
\eea
where $\rho_c=3 \mathcal{H}_0^2/8\pi G$ with $\mathcal{H}_0\simeq 2.2 \times 10^{-4}~\rm Mpc^{-1}$ being the Hubble constant. From Eq.~\eqref{gw1}, the quantity $\Omega_{\rm GW}(k)$ is derived as 
\bea
\Omega_{\rm GW}(k)=\frac{1}{12\mathcal{H}_0^2}\left(\frac{k}{a_0}\right)^2T_T^2(\tau_0,k)P_T(k)\,, ~~\tau_0=1.4\times 10^4 {\rm ~Mpc}\,.\label{GWeq}
\eea
The transfer function can be computed analytically, matching numerical results with a fair accuracy~\cite{t1,t2,t3,t4,t5,t6}. When an intermediate matter domination (recall that a RH neutrino mass-dependent intermediate matter domination is the central theme of this article) is at play, $T_T^2(\tau_0,k)$ can be calculated as~\cite{t5,t6}
\bea
T_T^2(\tau_0,k)=F(k) \,T_1^2(\zeta_{\rm eq}) \, T_2^2(\zeta_{\Phi}) \, T_3^2(\zeta_{\Phi R}) \, T_2^2(\zeta_{R})\,, \label{TT}
\eea
where the individual transfer functions read as
\begin{eqnarray}
    T_1^2(\zeta) &=& 1+1.57\,\zeta+ 3.42 \,\zeta^2\,,\\
    T_2^2(\zeta) &=& \left(1-0.22\,\zeta^{1.5}+0.65\,\zeta^2 \right)^{-1}\,,\\
    T_3^2(\zeta) &=& 1+0.59\,\zeta+0.65 \, \zeta^2\,,
\end{eqnarray}
with $\zeta_i =  k/k_i$. The modes $k_i$ given by
\begin{eqnarray}
    k_{\rm eq} &=& 7.1\times 10^{-2}\Omega_m h^2 {\rm Mpc^{-1}}\,\,\, {\rm with } \,\,\,
\Omega_m=0.31\,, \,\,\,h = 0.7\,,\\
    k_{\Phi} &=& 1.7\times 10^{14}\left(\frac{g_{*s}(T_{\rm dec})}{106.75}\right)^{1/6}\left(\frac{T_{\rm dec}}{10^7 \rm GeV}\right){\rm Mpc^{-1}} \,,\label{flow} \\
    k_{\Phi R} &=& 1.7\times 10^{14} \kappa^{2/3}\left(\frac{g_{*s}(T_{\rm dec})}{106.75}\right)^{1/6}\left(\frac{T_{\rm dec}}{10^7 \rm GeV}\right){\rm Mpc^{-1}}\,, \label{fdip}\\
    k_{R} &=&1.7\times 10^{14}\kappa^{-1/3}\left(\frac{g_{*s}(T_{R})}{106.75}\right)^{1/6}\left(\frac{T_{R}}{10^7 \rm GeV}\right){\rm Mpc^{-1}}\label{fhigh} \,,
\end{eqnarray}
re-enter the horizon at $T_{\rm eq}$ (the standard matter-radiation equality temperature), at $T_{\rm dec}$ (the temperature at which the scalar field decays), at $T_{\rm dom}$ (the temperature at which the scalar field starts to dominate) and at  $T_{R}$ (when the universe reheats after inflation), respectively. In our computation, we use the analytical expression for $\kappa$ presented in Eq.~\eqref{newentr}. Notice that we have to consider the reheating after inflation precedes the phase transition of $\Phi$, i.e., $T_{R}>T_c$ which does not contradict the condition for thermal leptogenesis $T_{R} > M_N$ because $T_c >M_N$ (see Eq.~\eqref{rhmass_b}). {\color{black} In the numerical analysis, we shall discuss the results for $T_c\lesssim T_R\lesssim T_R^{\max}$, where we consider $T_R^{\max}\simeq 10^{15}$ GeV as the maximum allowed reheating temperature.} The quantity $F(k)$ in Eq.~\eqref{TT} is given by
\bea
F(k) = \Omega_m^2\left( \frac{g_*(T_{k,\rm in})}{g_{*0}}\right)\left(\frac{g_{*0s}}{g_{*s}(T_{k,\rm in})}\right)^{4/3}\left(\frac{3j_1(k\tau_0)}{k\tau_0}\right)^2 \,, \label{fuk}
\eea
where $T_{k,\rm in}$ is the temperature corresponding to the horizon entry of the $k$th mode, $j_1(k\tau_0)$ is the spherical Bessel function, $g_{*0}=3.36$, $g_{*0s}=3.91$, and an approximate form of the scale-dependent $g_*$ is given by~\cite{gs1,gs2,t6}
\bea
g_{*(s)}(T_{k,\rm in})=g_{*0(s)}\left(\frac{A+{\tanh k_1}}{A+1}\right)\left(\frac{B+{\tanh k_2}}{B+1}\right)\,,
\eea
where 
\bea
A=\frac{-1-10.75/g_{*0(s)}}{-1+10.75/g_{*0(s)}}\,,~~B=\frac{-1-g_{\rm max}/10.75 }{-1+g_{\rm max}/10.75}\,, 
\eea
and
\bea
k_1=-2.5~\log_{10}\left(\frac{k/2\pi}{2.5\times 10^{-12}~{\rm Hz}}\right)\,, ~~k_2=-2.0~{\log}_{10}\left(\frac{k/2\pi}{6.0\times 10^{-9}~{\rm Hz}}\right)\,,
\eea
with $g_{\rm max}$ being $\simeq 106.75$. With the above set of equations, we evaluate the normalized GW energy density $\Omega_{\rm GW}(k)$ in Eq.~\eqref{GWeq} for the benchmark parameters considered in the previous section and reported in Tab.~\ref{tab:BP}. Before we present the numerical results, let us point out that the quantity $\Omega_{\rm GW}(k)h^2$ is constrained by two robust bounds on the Stochastic Gravitational Wave Background (SGWB) placed by the effective number of relativistic species during Big Bang Nucleosynthesis (BBN)~\cite{Peimbert:2016bdg} and by the LIGO measurements~\cite{LIGOScientific:2016jlg}. In particular, the BBN constraint reads
\bea
\int_{f_{\rm low}}^{f_{\rm high}} {\rm d} f ~f^{-1}\Omega_{\rm GW}(f)h^2\lesssim 5.6\times 10^{-6}\Delta N_{\rm eff}\,,
\eea
with $\Delta N_{\rm eff}\lesssim 0.2$. The lower limit of the integration is the frequency that represents the mode entering the horizon at the BBN epoch, which we take as $f_{\rm low}\simeq 10^{-10}$ Hz. {\color{black} We consider the upper limit $f_{\rm high}\simeq 10^{7}$ Hz corresponding to $T_R\equiv T_R^{\rm max}\simeq 10^{15}$ GeV. For the LIGO constraint, we rely on a crude estimation by considering a reference frequency $f_{\rm LIGO}=25~{\rm Hz}$ and discarding the GWs having an amplitude more than $8.33\times 10^{-9}$ at $f_{\rm LIGO}$~\cite{KAGRA:2021kbb}. On the other hand, in case of currently unconstrained GW signals, we quantitatively assess their detectability by next-generation detectors after $t_{\rm obs} = 1~{\rm yr}$ observing time by computing the signal-to-noise ratio (SNR) defined as~\cite{Maggiore:1999vm,Schmitz:2020syl}
\bea
{\rm SNR} = \sqrt{t_{\rm obs}\int_{f_{\rm min}}^{f_{\rm max}}df \left(\frac{\Omega_{\rm GW}(f)}{\Omega_{\rm noise}(f)}\right)^2}\,,\label{snr}
\eea
where $\Omega_{\rm noise}(f)$ is the noise spectrum of the detector and $f_{\rm min}(f_{\rm max})$ are minimum (maximum) accessible frequency.}

\section{Results}\label{s4}

\begin{figure}[t!]
\includegraphics[width=0.49\textwidth]{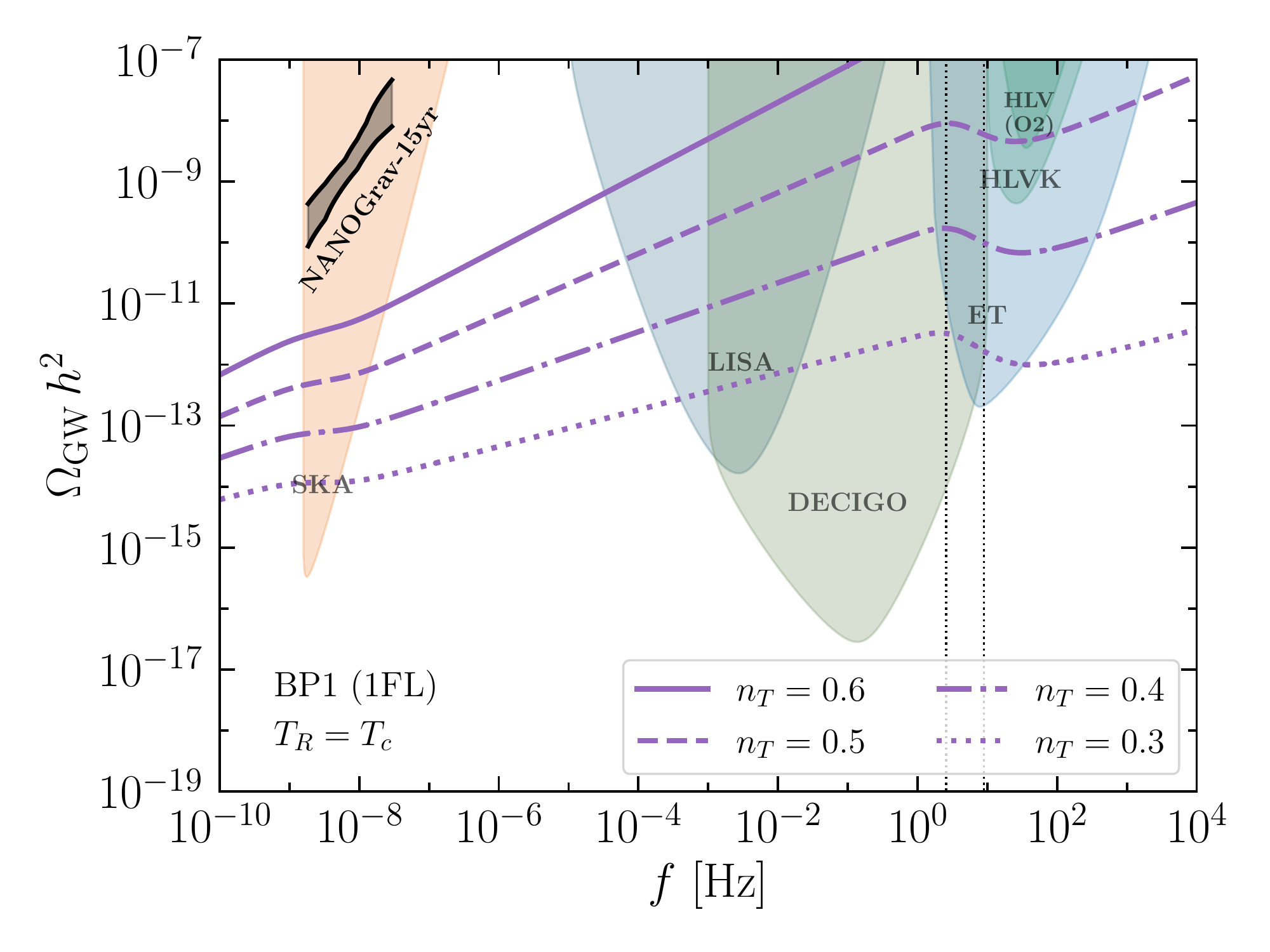}
\includegraphics[width=0.49\textwidth]{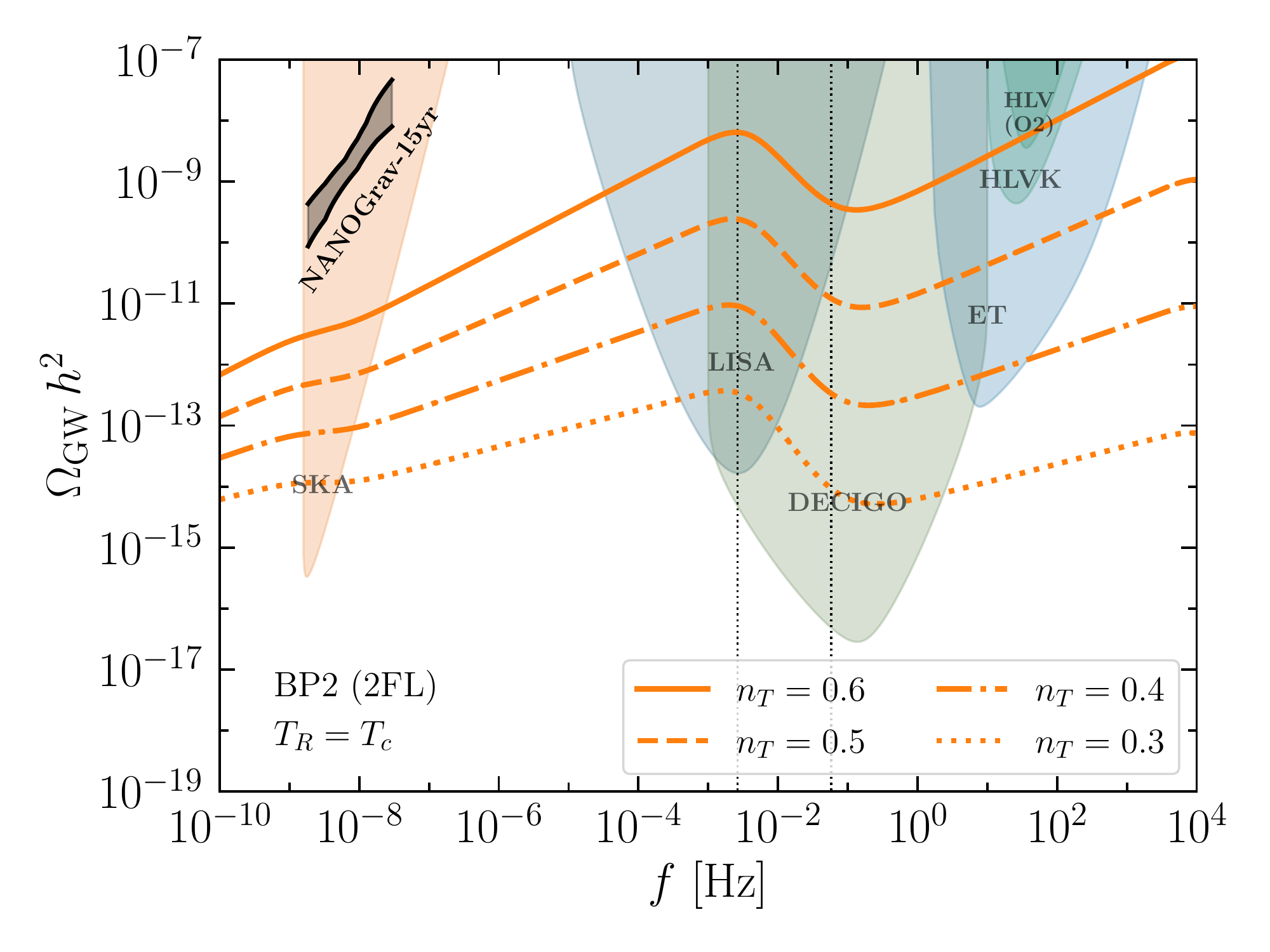}
\includegraphics[width=0.49\textwidth]{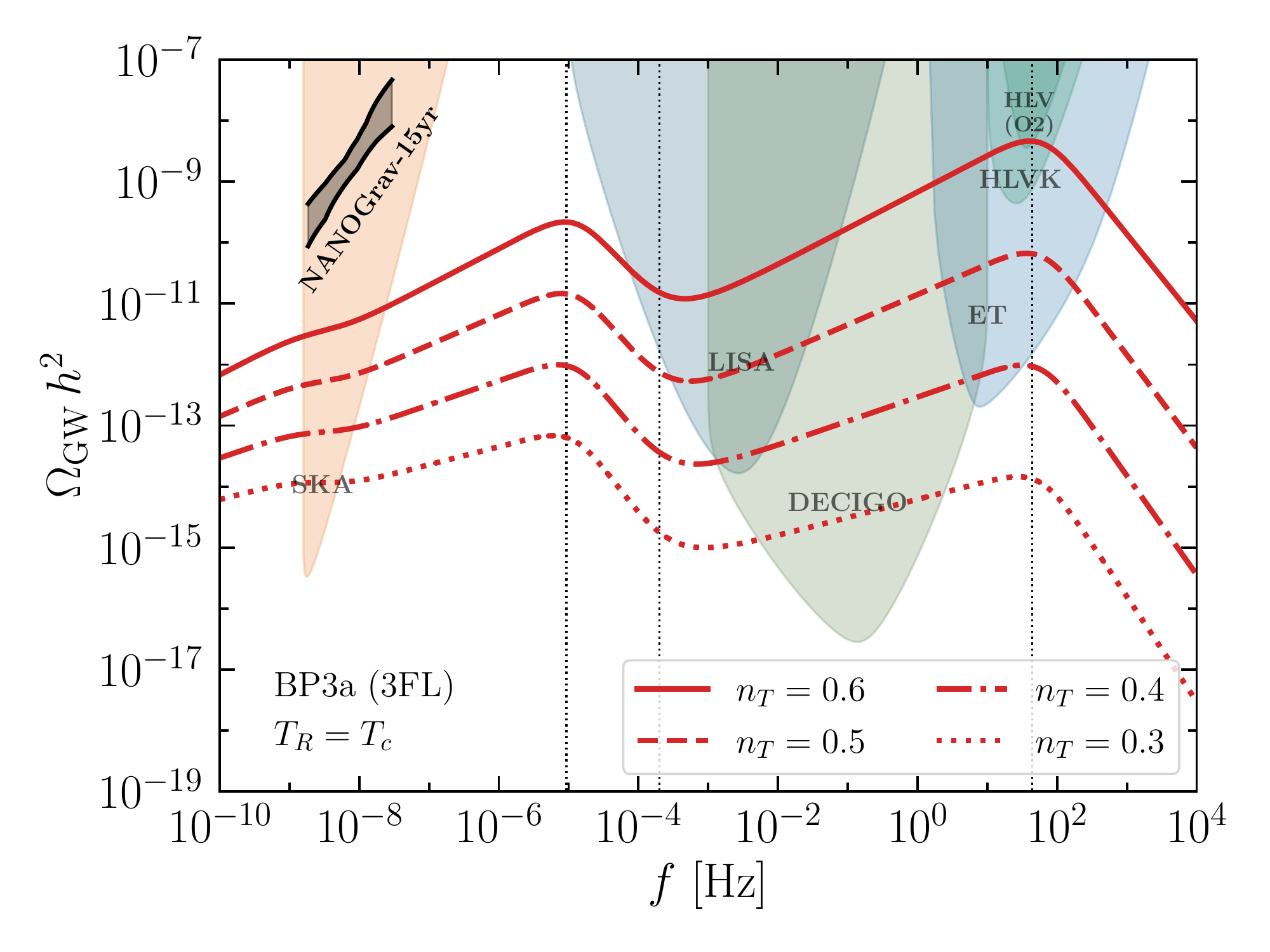}
\includegraphics[width=0.49\textwidth]{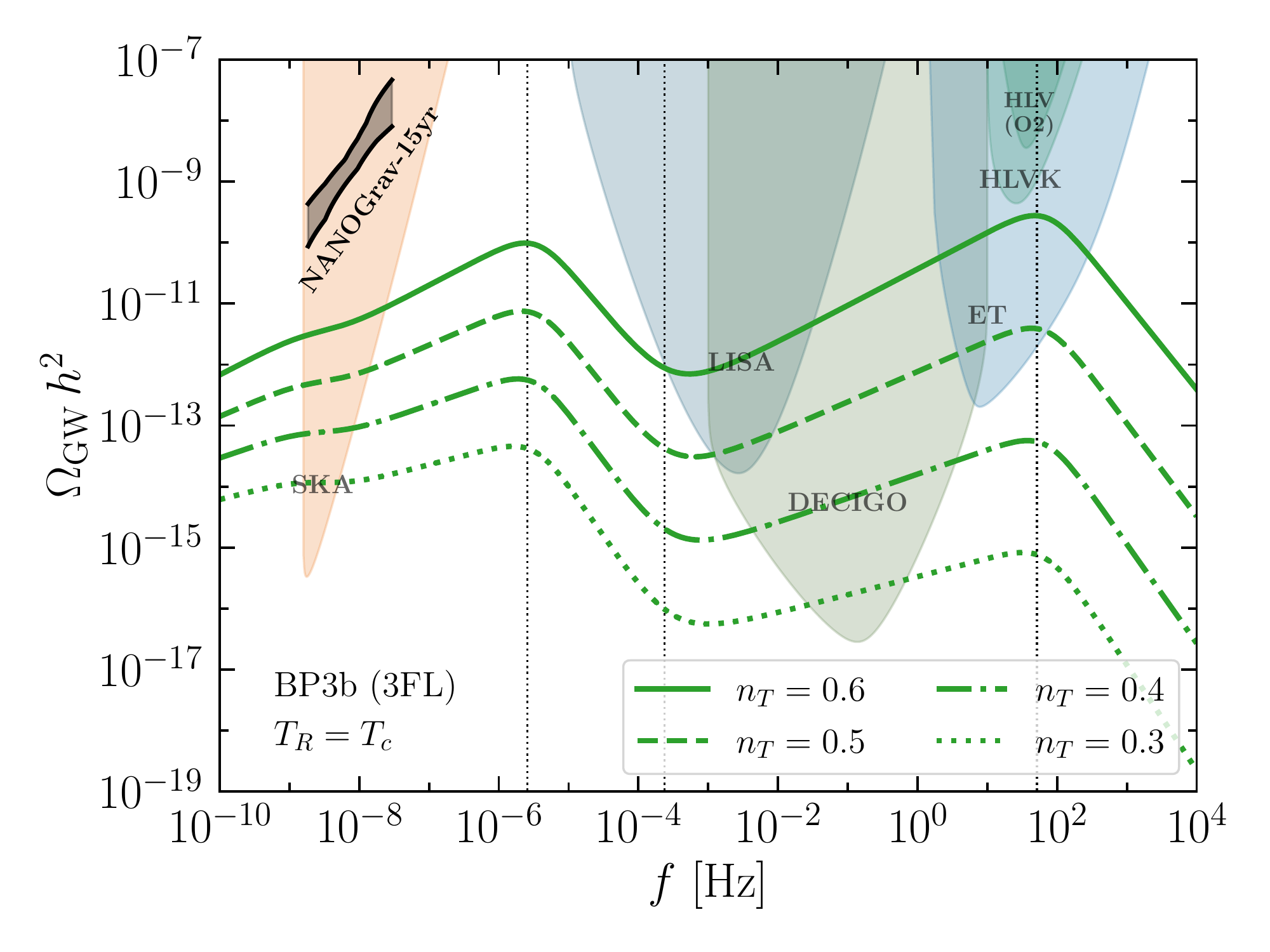}
\caption{BGW spectra in case of four different values of the tensor spectral index $n_T$, assuming $T_R = T_c$. The different plots correspond to the four benchmark points (given in Tab.~\ref{tab:BP}) depicting different entropy production and flavour regimes of leptogenesis. The shaded regions
are sensitivity curves of various GW experiments such as SKA~\cite{ska}, LISA~\cite{Baker:2019nia}, DECIGO~\cite{Isoyama:2018rjb}, ET~\cite{Maggiore:2019uih}, and HLVK~\cite{LIGOScientific:2019hgc}. The black bands in the nHz region represents the NANOGrav 15yrs data posteriors~\cite{ng1}. The dotted vertical lines (from left to right) correspond
to $f_{\rm low}^{\rm peak}$, $f^{\rm dip}$, and $f_{\rm high}^{\rm peak}$, respectively.}
\label{fig:GW} 
\end{figure}
\begin{figure}[t!]
\includegraphics[width=0.49\textwidth]{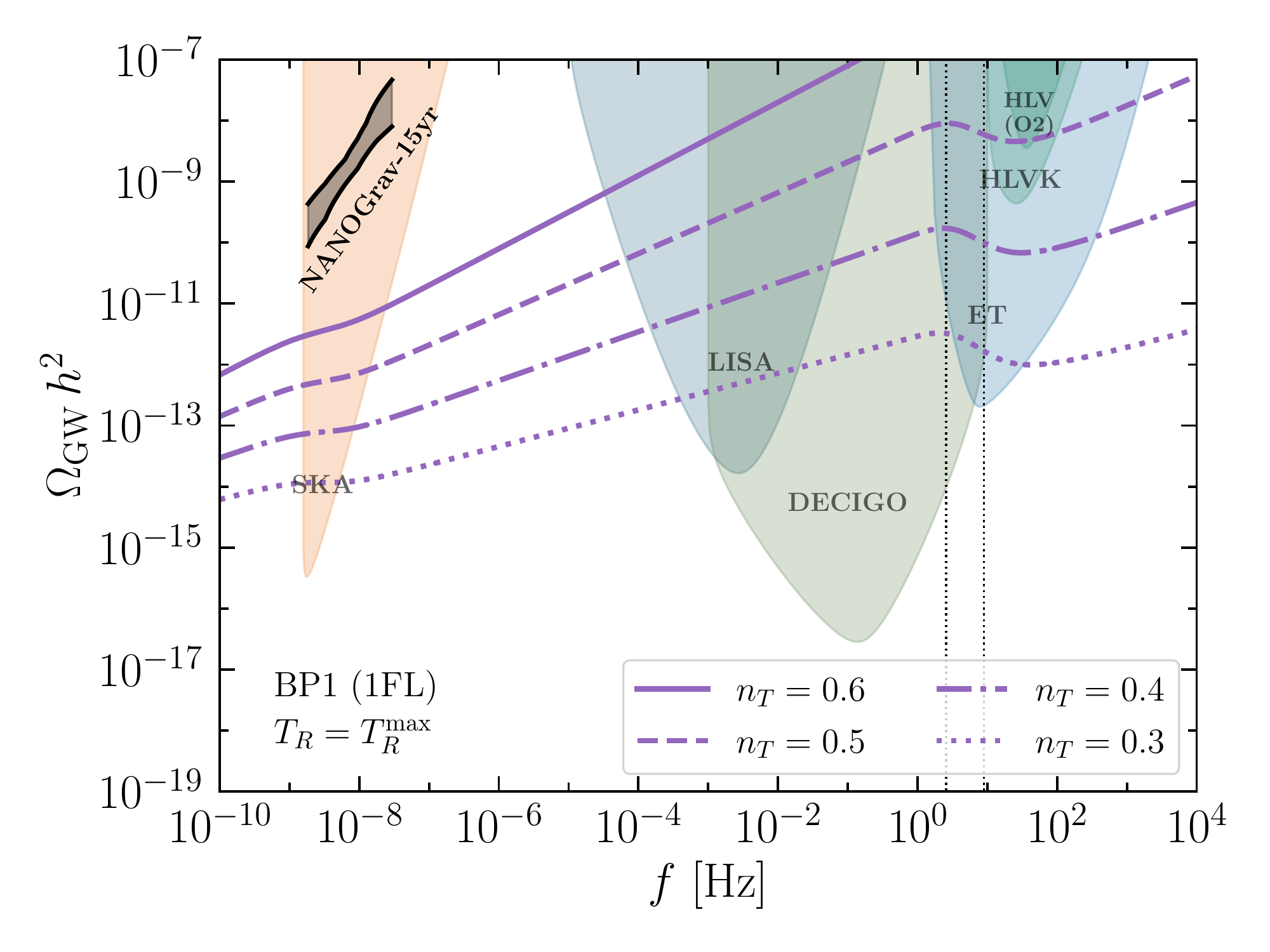}
\includegraphics[width=0.49\textwidth]{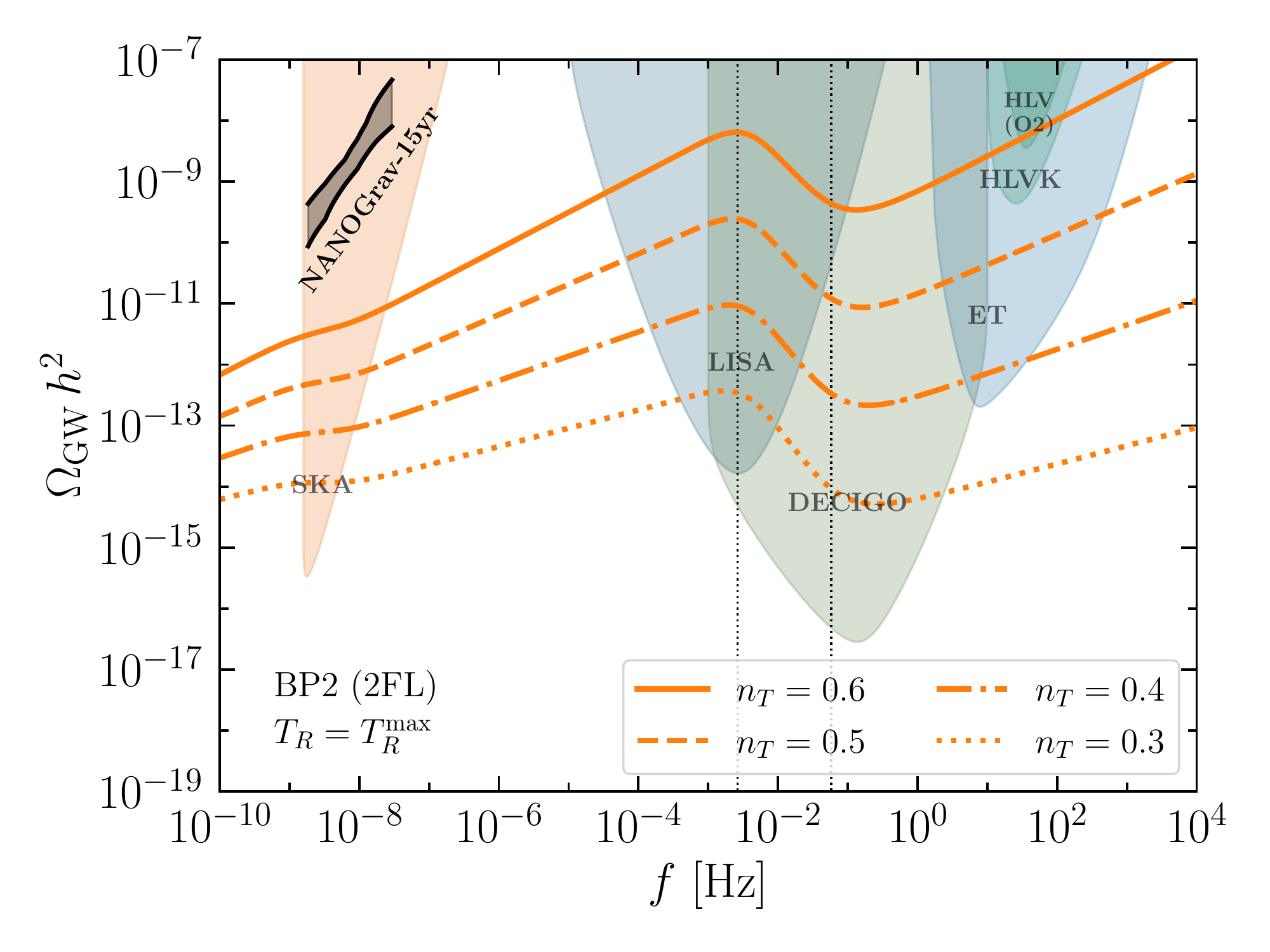}
\includegraphics[width=0.49\textwidth]{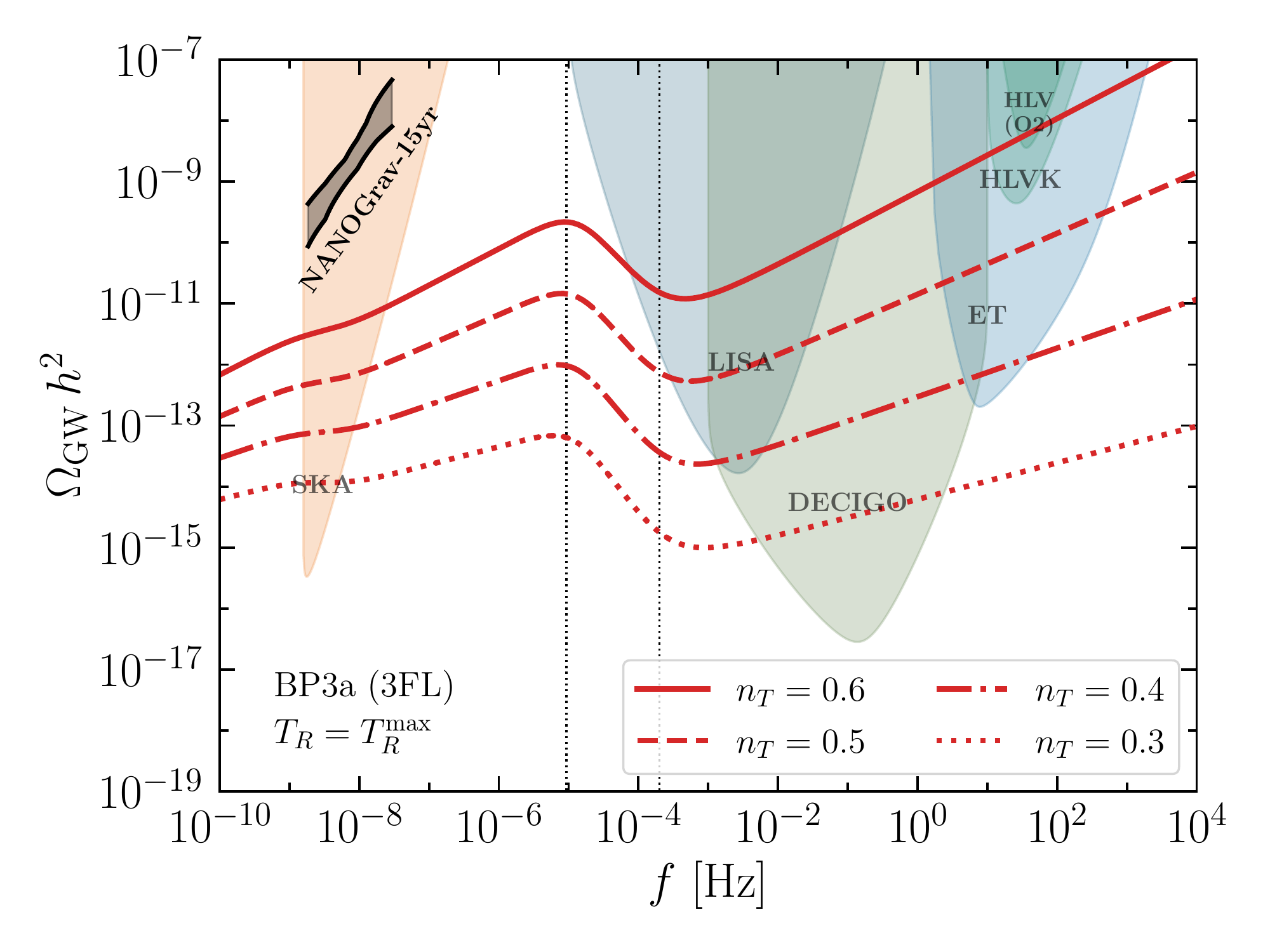}
\includegraphics[width=0.49\textwidth]{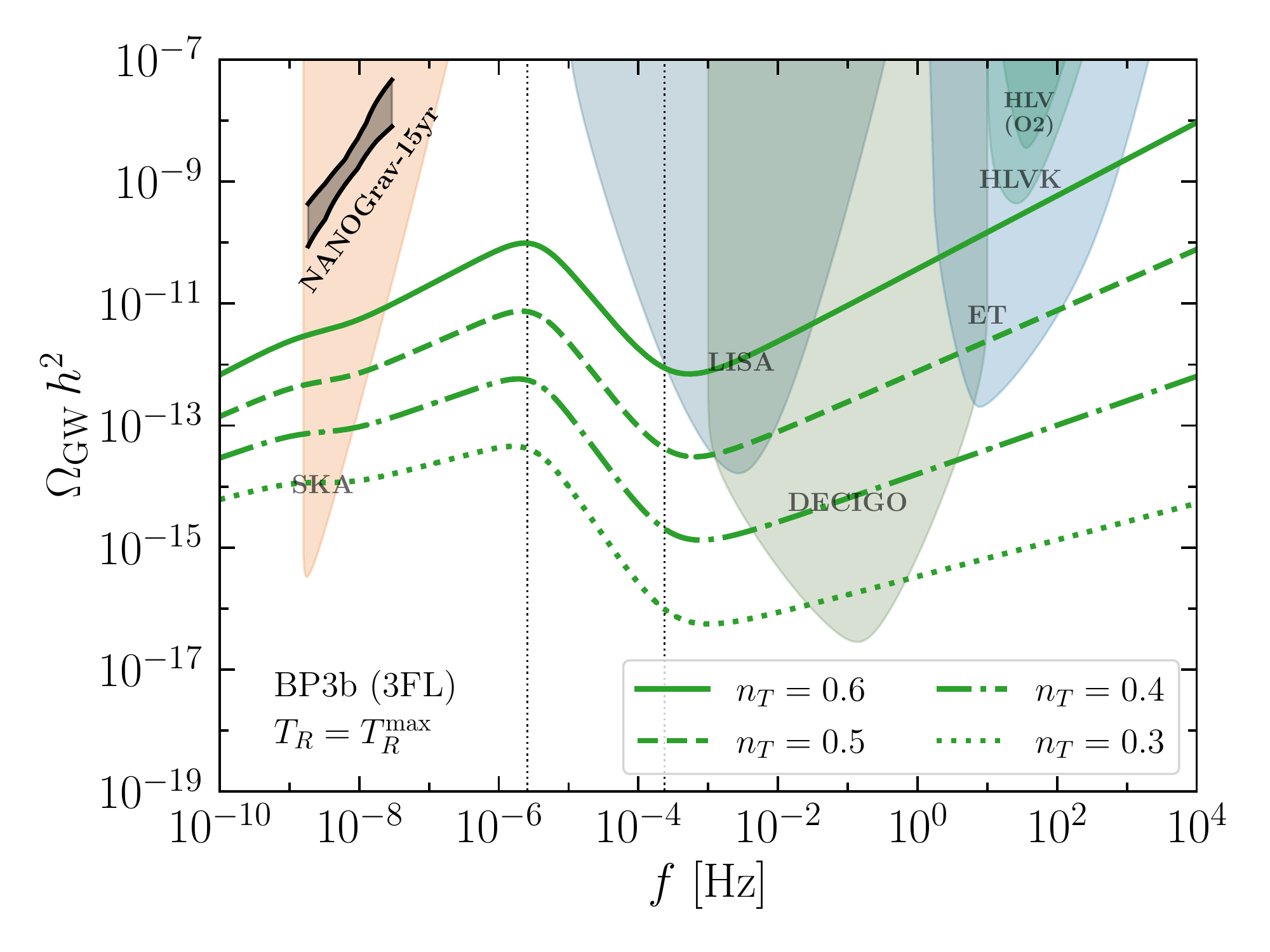}
\caption{Same as Fig.~\ref{fig:GW} but assuming $T_R = T_R^{\rm max} = 10^{15}~{\rm GeV}$. For all the GW spectra, the second peak frequency $f_{\rm high}^{\rm peak}$ occurs at $10^7~{\rm Hz}$, which is beyond the displayed range.}
\label{fig:GWmax} 
\end{figure}

{\color{black} In this section, we present our numerical results on the spectral features in the BGW spectra and quantitatively assess their potential detection in future experiments. We consider the case of $g^\prime = 10^{-2}$ and discuss the impact of the different values for the gauge coupling in the Appendix~\ref{app:1}, where we demonstrate that our results are qualitatively unchanged.}

{\color{black} In Fig.s~\ref{fig:GW} and~\ref{fig:GWmax}, we present the normalized GW energy density as a function of the frequency and the spectral features of the BGWs induced by the $\Phi$ matter domination set by the RH neutrino mass scale for different values of tensor spectral index $n_T$. These figures differ from the assumption on the reheating temperature $T_R$ which is taken to be $T_c$ and $T_R^{\rm max} = 10^{15}~{\rm GeV}$, respectively. In both figures, the different panels correspond to the four leptogenesis benchmark points presented in Tab.~\ref{tab:BP}. The GW spectra are characterized by three frequencies, $f_{\rm low}^{\rm peak}$, $f^{\rm dip}$, and $f_{\rm high}^{\rm peak}$, which correspond to the temperatures $T_{\rm dec}$, $T_{\rm dom}$, and $T_{R}$ (cf Eq.s~\eqref{flow},~\eqref{fdip} and~\eqref{fhigh}). Notice that irrespective of the spectral index $n_T$, the overall spectrum gets more suppressed for large entropy production. In addition, the red tilt in the middle becomes more prominent because longer duration of $\Phi$ domination delays the horizon entry of scales $k^{-1}\in \left[k_\Phi^{-1},k_{\Phi R}^{-1}\right]$, and the corresponding amplitude within these $k$ values get reduced. Therefore, as the RH neutrino mass scale decreases (e.g. going from BP1 to BP3), we get a more distinct double-peaked spectrum in case of $T_R = T_c$. As such, a three-flavour leptogenesis scenario (BP3b: bottom-right panel) can predict a more distinct double peak spectrum than a two-flavour (BP2: top-right panel) or one-flavour (BP1: top-left panel) scenario. Note also that though the spectral distortion, i.e., the ratio $\Omega_{\rm GW}(f_{\rm low}^{\rm peak})/\Omega_{\rm GW}(f^{\rm dip})$ is same for BP2 and BP3a, the peak and the dip frequencies are in different position as $T_{\rm dec(dom)}$ is different for these benchmark points.  In addition to that, starting from a one-flavour/vanilla regime, as a leptogenesis scenario goes deep into the flavour regime, the low-frequency peak and the dip shift towards lower frequencies. As evident from Fig.~\ref{fig:GW}, these frequencies (two-flavour scenario) fall within the LISA-DECIGO (mHz-Hz) sensitivity \cite{lisa,decigo}, whereas, for BP3 (three-flavour scenario), it falls in the mHz-$\mu$Hz region.}
\begin{figure}[t!]
\includegraphics[width=\textwidth]{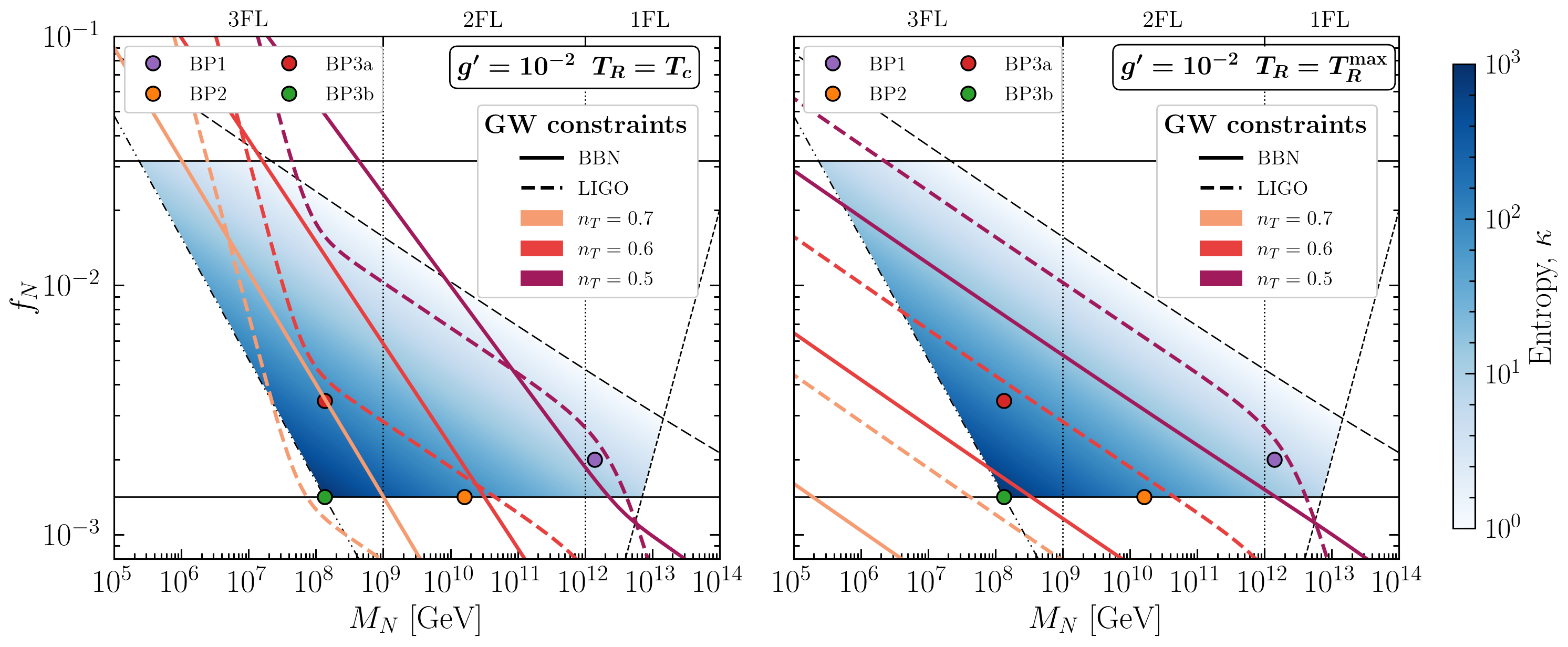}
\caption{BBN (solid) and LIGO (dashed) constraints on the $M_N$-$f_N$ plane for different values of the tensor spectral index $n_T$ in case of $T_R = T_c$ (left panel) and $T_R = T_R^{\rm max}$ (right panel). Anything to the right of these lines is excluded. Note that the constraints (especially the BBN one) depend on the reheating temperature $T_R$, implying that a viable tomography of flavoured leptogenesis with BGWs can be done only for $n_T \lesssim 0.7$.}
\label{fig:GWconstraints} 
\end{figure}

{\color{black} In Fig.~\ref{fig:GWconstraints}, we translate the BBN (solid lines) and LIGO (dashed lines) constraints on $\Omega_{\rm GW}$ into bounds on the model parameters and, remarkably, on the RH neutrino mass scale. These constraints depend on the value of the tensor spectral index $n_T$ (different colored lines) and on the reheating temperature of the universe for which we consider the two opposite cases: $T_R = T_c$ (left panel) and $T_R = T_R^{\rm max} = 10^{15}~{\rm GeV}$ (right panel). Notice that the parameter space to the right of the displayed lines is excluded. In case of $T_R = T_c$, we find the most conservative limits and the LIGO constraint to be in general much stronger than BBN one, ruling out almost the entire parameter space under consideration for $n_T = 0.7$.  On the other hand, the case $T_R = T_R^{\rm max}$ correspond to the strongest limits on the model parameter space, which is completely ruled out for $n_T = 0.6$. The allowed parameter space opens for smaller values of $n_T$. We find that the three-flavour and two-flavour regimes are allowed in case of $n_T\lesssim0.6$ and $n_T \lesssim 0.5$ for $T_R = T_c$ and $T_R = T_R^{\rm max}$, respectively. For $n_T \ll 0.5$, most of the model parameter space evades instead all the GW bounds. An important aspect of the obtained GW spectrum is that the overall amplitude increases with $n_T$, and the spectrum spans a wide range of frequencies. Therefore, one way to predict robustly testable GW signals is to fix a UV model that would predict a particular value of $n_T$ ($\ll 0.7$ for the flavoured leptogenesis model as discussed here), which is not our approach. Another way, which we discuss here, is to take an IR approach by being agnostic about the source of the BGWs and treating $n_T$ as a free parameter. In this case, if we are able to fit a GW signal within a certain frequency band for a particular range of $n_T$ values, we can robustly predict the GW spectrum with distinct spectral features testable at a different frequency band. As such, as shown in Fig.s~\ref{fig:GW} and~\ref{fig:GWmax} (BP3a, bottom-left panels), a three-flavour leptogenesis scenario can predict a testable signal in the HLVK band for $0.5 \lesssim n_T\lesssim 0.6$. Therefore, if the future HLVK run finds a stochastic GW signal aligned with our anticipation, a three-flavour leptogenesis scenario would predict a signal with the similar spectral slope in the LISA range, distinct spectral behaviour in the $\mu$Hz range plus a testable blue tilted signal at the nHz frequencies. We can in fact compare this signal in the nHz band with the recent PTA data. For this, we first opt for the PTA parameterization for the $\Omega_{\rm GW}(f)$, which reads: 
\begin{equation}
\Omega_{\rm GW}(f)=\Omega_{\rm yr}\left(\frac{f}{f_{\rm yr}}\right)^{(5-\gamma)}\,,\label{pl}
\end{equation}
where $\Omega_{\rm yr}=2\pi^2 A^2 f_{\rm yr}^2 /(3H_0^2)$, $f_{\rm yr}=1~{\rm yr^{-1}}\simeq 32~{\rm nHz}$ with $A$ and $\gamma$ being the amplitude and spectral index having best-fit values\footnote{Here we quote the NANOGrav best-fit values. Nonetheless, recently reported global posteriors \cite{ipta} that include results from all the PTAs are similar to NANOGrav, because NANOGrav offers more statistically significant data among all the PTAs.}  $A\sim 6.4_{-2.7}^{+4.2}\times 10^{-15}$ and $\gamma \sim 3.2\pm 0.6$. In the PTA frequency band, the $\Omega_{\rm GW}(f)$ in Eq.~\eqref{GWeq} is basically $\Omega_{\rm GW}(f)\propto f^{n_T}$, because in this range, $T_1(\xi)$ dominates, and we can approximate $j_1(k\tau_0)\simeq 1/k\tau_0$. As a result, the power spectrum $P_T(k)$ in Eq.~\eqref{ptp} determines the overall spectral shape.  We can therefore compare Eq.~\eqref{pl} and Eq.~\eqref{GWeq} to extract $A$ and $\gamma$. For example,  BP3a ($n_T=0.6$) would provide $A\simeq 1.2\times 10^{-16}$ and $\gamma \simeq 5-n_T=4.4$. Although the spectral index can be reconciled with the PTAs at 3$\sigma$ CL, the amplitude is off by an order of magnitude.  Remarkably, in the PTA band, this signal however mimics the one expected from GW-driven SMBHB models~\cite{Phinney:2001di} which predict $\gamma=13/3$ and with $ A^{\rm SMBHB}$ in the $10^{-16}$ ballpark. As discussed, with our BP3a a fit to an observation of SGWB by HLVK would imply a robustly testable signal in the nHz range with $A\sim A^{\rm SMBHB}$. Therefore, if the recent PTA finding persists, then future non-observation of nHz GWs with $A\sim A^{\rm SMBHB}$ would imply that BP3a (or, more generally, a part of our model parameter space) would be ruled out. Notice that, the GW spectrum around the spectral index $n_T\simeq 0.6$ is the strongest and perhaps the most promising signal that the discussed flavoured leptogenesis scenario offers. This is because, in this ballpark, GWs, apart from having amplitudes $A\sim A^{\rm SMBHB}$ at nHz and $A\sim A^{\rm HLVK}$  at Hz, exhibit prominent and testable spectral features in the $\mu$Hz and mHz bands. A higher value of $n_T$ saturates LIGO and BBN  bounds as shown in Fig.~\ref{fig:GWconstraints} (this is the reason that our scenario never produces a signal with $A\sim A^{\rm PTAs}$). On the contrary, for $n_T\ll 0.6$, although the overall signal strength decreases and the spectrum in the nHz range loses testability, at higher frequencies the model can still provide signals with reasonably high SNR for spectral index as low as $n_T\simeq 0.3$. Below we discuss this by performing a rigorous parameter space scan of the model and taking into account all the constraints that the model has to comply with. }

\begin{figure}[t!]
\includegraphics[width=\textwidth]{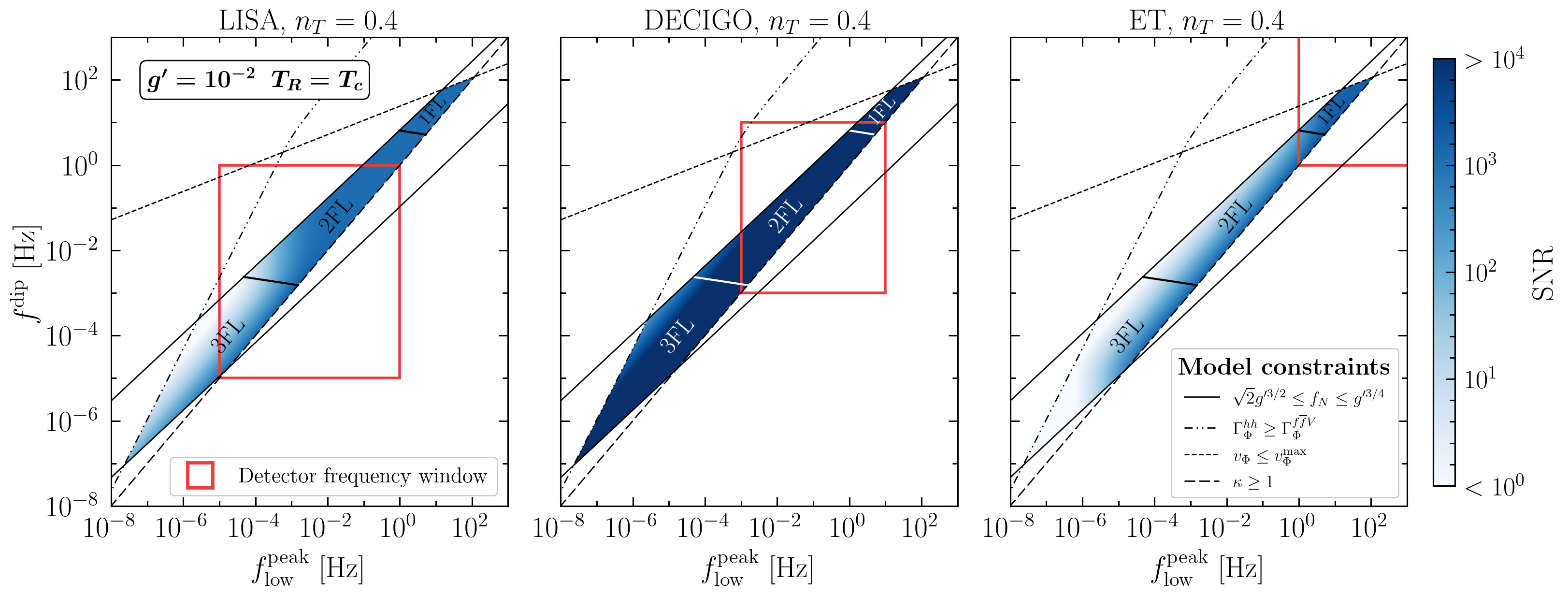}\\
\vspace{0.3cm}
\includegraphics[width=\textwidth]{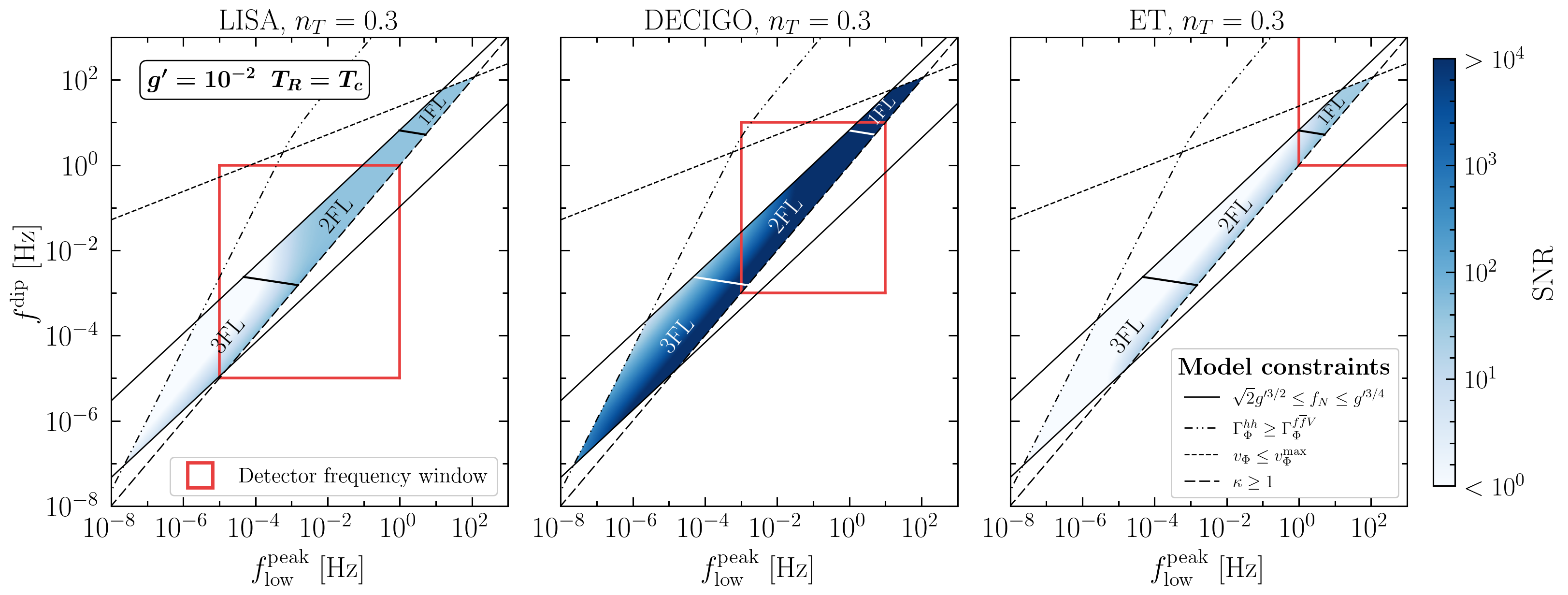}\\
\caption{Signal-to-noise ratio (SNR) with 1-year observing time (see Eq.~\eqref{snr}) for LISA, DECIGO and ET detectors in case of a GW signal with $n_T = 0.4$ (upper panels) and $n_T = 0.3$ (lower panels) in the $f_{\rm low}^{\rm peak}$-$f^{\rm dip}$ plane. The two frequencies are uniquely determined by the model parameters $M_N$ and $f_N$, and they are bounded by the thin black lines defined by several model requirements (see also Fig.~\ref{fig:model}). The red squares highlight the frequency sensitivity windows of each detector, which can be compared with the regions of the parameter space achieving three-flavour (3FL), two-flavour (2FL) and one-flavour (1FL) leptogenesis. The detectors will observe the GW spectral distortions induced by the model parameters inside the red squares.}
\label{fig:SNR}
\end{figure}

{\color{black} In Fig.~\ref{fig:SNR}, we report our SNR computation according to Eq.~\eqref{snr} in order to quantify the detectability of the GW signal with $n_T = 0.4$ (upper panels) and $n_T = 0.3$ (lower panels), which evade all the present constraints. In each row, the three plots from left to right correspond to the future experiments LISA, DECIGO and ET, respectively. In each plot, the thin black lines delimit the allowed parameter space in the plane according to the model constraints previously discussed. Notice that the model parameters $M_N$ and $f_N$ uniquely determine the spectral features of the GW signal characterized by the low peak frequency $f_{\rm low}^{\rm peak}$ and the dip frequency $f^{\rm dip}$. Remarkably, as can be seen from the plots, the different ranges for $M_N$ related to the different flavour regimes of leptogenesis are separated with 3FL, 2FL and 1FL predicting frequencies in the nHz-$\mu$Hz, sub-Hz and Hz ranges, respectively. This further confirms that the spectral features of the GW signals trace the leptogenesis flavour regimes. We find that all the three GW detectors will potentially observe a GW signal with $n_T \gtrsim 0.3$ having ${\rm SNR \geq 1}$ in most of the parameter space, while we check that DECIGO can probe GW signals with $n_T \gtrsim 0.1$. The ET detector and partially the LISA one show a smaller sensitivity to the 3FL parameter space. This trend is indeed expected since the 3FL parameters cause a distortion of the GW signal at sub-$\mu$Hz frequencies and, consequently, lower the GW amplitude at higher frequencies hampering the detection. We stress that, in general, the SNR does not carry any spectral information being dependent on the integrated signal (see Eq.~\eqref{snr}) within the detector frequency intervals, which are highlighted with red squares in the plots. However, inside the red squares, the detectors would actually observe the spectral distortions ($f_{\rm low}^{\rm peak}$ as well as $f^{\rm dip}$) and not just a GW power-law signal. Therefore, we find that LISA, DECIGO and ET detectors will robustly probe the three-flavour (3FL), two-flavour (2FL) and one-flavour (1FL) regimes of leptogenesis, respectively. We emphasize that our results are not considerably affected by the choice of the reheating temperature, which is assumed to be $T_R = T_c$ in Fig.~\ref{fig:SNR}. On the contrary, the higher the reheating temperature, the higher the GW amplitude at frequencies higher than Hz, thus increasing the SNR at the ET detector while leaving unchanged the SNR at the LISA and DECIGO detectors.} 

The following concluding remarks are in order. First, a generic feature of $\Omega_{\rm GW}\propto f^{n_T\gg 0}$ is that the spectrum does not span a wide range of frequencies due to the BBN bound on effective neutrino species.  Therefore, cosmologically viable inflationary BGWs require a low reheating temperature after inflation. This has been discussed in detail in the literature and recently in the context of PTA results \cite{bn5n}. However, when an intermediate matter-dominated epoch is at play, the spectrum may span a wide range of frequencies without violating any cosmological bounds because of the entropy injection. It allows a high $T_R$, which is notably, also a natural requirement for high-scale thermal leptogenesis, as we discuss here. Therefore, high-scale leptogenesis scenarios are ideal to be searched for with BGW-spectroscopy. {\color{black} Let us also stress that our results depend on the parameterization of the primordial power spectrum in Eq.~\eqref{ptp} and the constant value of $n_T$. A detailed study with a different parameterization \cite{t6,Jiang:2023gfe,Cai:2020qpu,Giare:2022wxq} plus considering a running of $n_T$~\cite{Kuroyanagi:2011iw,Calcagni:2020tvw} which might induce a possible degeneracy in the predicted GW signals, is beyond the scope of this article.} Second, since the last couple of years, there has been an effort to test leptogenesis with primordial GWs, see, e.g., Refs.~\cite{Blasi:2020wpy,lepgw1,lepgw2,lepgw3,lepgw4,lepgw5,lepgw6,lepgw7,Ghoshal:2022kqp,lepgw8,Borah:2022cdx,Chun:2023ezg,Azatov:2021irb,Barman:2022gjo}. This article, nonetheless, for the first time, attempts to probe flavour regimes of leptogenesis with GWs. Finally, because we consider a $U(1)$ gauge symmetry breaking, our scenario must produce cosmic strings \cite{Kibble:1976sj}, which is another source of GWs \cite{Vilenkin:1981bx}. We nonetheless note that we work with unconventional parameter space, which in the numerical simulations of the cosmic strings evolution are taken as $g^\prime \sim ~\mathcal{O}(1)$ and $\lambda \sim ~\mathcal{O}(1)$ \cite{Blanco-Pillado:2011egf,Matsunami:2019fss}. A semi-analytical description covering this topic can be found in Ref.\cite{Chianese:2024gee} which assumes that the standard computation of GW spectrum from cosmic strings holds for smaller values of $g^\prime$ and $\lambda$. In this article, we restrict the GW signal's detectability at the SNR level. In a follow-up publication we shall present a more precise statistical analysis to properly quantify the level of uncertainties in the reconstruction of leptogenesis parameters by means of a Fisher matrix analysis and a Markov Chain Monte Carlo estimation~\cite{Flauger:2020qyi,f1,Giese:2021dnw,f2}.

\section{Summary}\label{s5}

In the present paper, for the first time, we show that flavour regimes of thermal leptogenesis can be markedly tested with gravitational waves' spectral features. Flavour effects bridge high-scale leptogenesis and low-energy neutrino physics and are important in leptogenesis computation. We opt for a tomographic method, where gravitational waves originating from an independent source, confront the leptogenesis model during their propagation in the early universe and carry imprints of leptogenesis parameters on the final gravitational waves spectrum. There are three distinct regimes of flavoured leptogenesis characterized by the right-handed neutrino mass scale: $M_N\gtrsim 10^{12}~{\rm GeV}$ (one-flavour/vanilla), $10^9~{\rm GeV}\lesssim M_N\lesssim 10^{12}~{\rm GeV}$ (two-flavour), and  $M_N\lesssim 10^{9}~{\rm GeV}$ (three-flavour). Testing flavour regimes with gravitational waves is thus equivalent to studying detectable spectral features in the gravitational waves dependent on the mass scale $M_N$. This is possible in the simplest and well-motivated realizations of seesaw scenarios that facilitate a scalar field to give mass to the right-handed neutrinos. Apart from generating right-handed neutrino masses, the scalar field can be long-lived to provide a matter-dominated epoch with a lifetime depending on $M_N$. When gravitational waves pass through such a matter-dominated phase, the start, the end, and the duration of the matter-domination get imprinted on the final gravitational waves spectrum. Thus in our case, the spectral features in the GWs become dependent on $M_N$, hence on different regimes of leptogenesis. While any gravitational waves that originate prior to the scalar field domination can offer a viable tomography, here we consider inflationary blue gravitational waves, which are now being widely discussed after the recent finding of blue stochastic gravitational waves by the Pulsar Timing Arrays, with an infra-red tail characterized by a simple power-law $\Omega_{\rm GW}\sim f^{n_T}$. We show that owing to the scalar field domination, the final gravitational waves exhibit a double peak spectrum, with the characteristic frequencies (the low-frequency peak and the dip between two peaks) depending on $M_N$. As the leptogenesis process enters deep into the flavour regimes, the spectral features in the gravitational waves become more prominent, and the characteristic frequencies shift to lower values (see Fig.s~\ref{fig:GW} and ~\ref{fig:GWmax}). As such, for a two-flavour (three-flavour) regime, the low-frequency peak shows up in the mHz ($\mu$Hz) bands.  While the BBN constraint on the effective number of neutrino species and LIGO bound on stochastic gravitational waves background restrict the spectral index value $n_T\lesssim 0.7$  (see Fig.~\ref{fig:GWconstraints}), a three-flavour leptogenesis scenario offers the most promising signal for $n_T \sim 0.5-0.6$. Indeed, it shows an amplitude testable in the next LIGO run at $f\sim $ Hz, characteristic spectral features in the mHz and $\mu$Hz bands, plus a refutable gravitational wave strain at $f\sim $ nHz, comparable to the one expected from supermassive black holes. {\color{black} Even in case of a smaller tensor spectral index, future detectors such as LISA, DECIGO and ET will be able to probe the spectral distortions connected to the scale $M_N$ of leptogenesis and therefore will provide remarkable insight into the leptogenesis flavour regimes (see Fig.~\ref{fig:SNR}).}

\section*{Acknowledgements}

The work of MC, RS and NS is supported by the research project TAsP (Theoretical Astroparticle Physics) funded by the Istituto Nazionale di Fisica Nucleare (INFN). The work of NS is further supported by the research grant number 2022E2J4RK ``PANTHEON: Perspectives in Astroparticle and Neutrino THEory with Old and New messengers'' under the program PRIN 2022 funded by the Italian Ministero dell’Università e della Ricerca (MUR). The work of SD is  supported by the National Natural Science Foundation of China (NNSFC) under grant No. 12150610460.

\appendix
\section{Varying the $B-L$ gauge coupling \label{app:1}}

Similar to Fig.~\ref{fig:GWconstraints} ($g^\prime=10^{-2}$), to show the qualitative trend and dependence on gauge coupling, we present the allowed parameter space for $g^\prime=0.05$ and $10^{-3}$ in Fig.s~\ref{fig:GWconstraints1} and~\ref{fig:GWconstraints3}, respectively. Note that larger values of gauge coupling result in a shrinking of the parameter space because of the imposed constraints and lesser entropy production (cf. Eq.~\eqref{rhmass_b}, Eq.~\eqref{newgamma}, and Eq.~\eqref{newentr}). As discussed in the main text, the smaller the entropy production, the less the suppression in GWs. Therefore, compared to the  $g^\prime=10^{-2}$  case, smaller values of the spectral indices are now allowed by the LIGO and the BBN. On the contrary, for smaller values of $g^\prime$, the allowed parameter space enlarges plus the LIGO and the BBN limits allow for larger values of the spectral indices. Note in either case, the qualitative features in the GW spectrum, however, remain intact. For example, for a 3FL we would get the low-frequency peak at smaller frequencies than a 2FL. Compared to the $g^\prime=10^{-2}$ case, however, this peak would shift to the right (left) for larger (smaller) values of gauge couplings (cf. Fig.~\ref{fig:GW}).

\begin{figure}[h!]
\includegraphics[width=\textwidth]{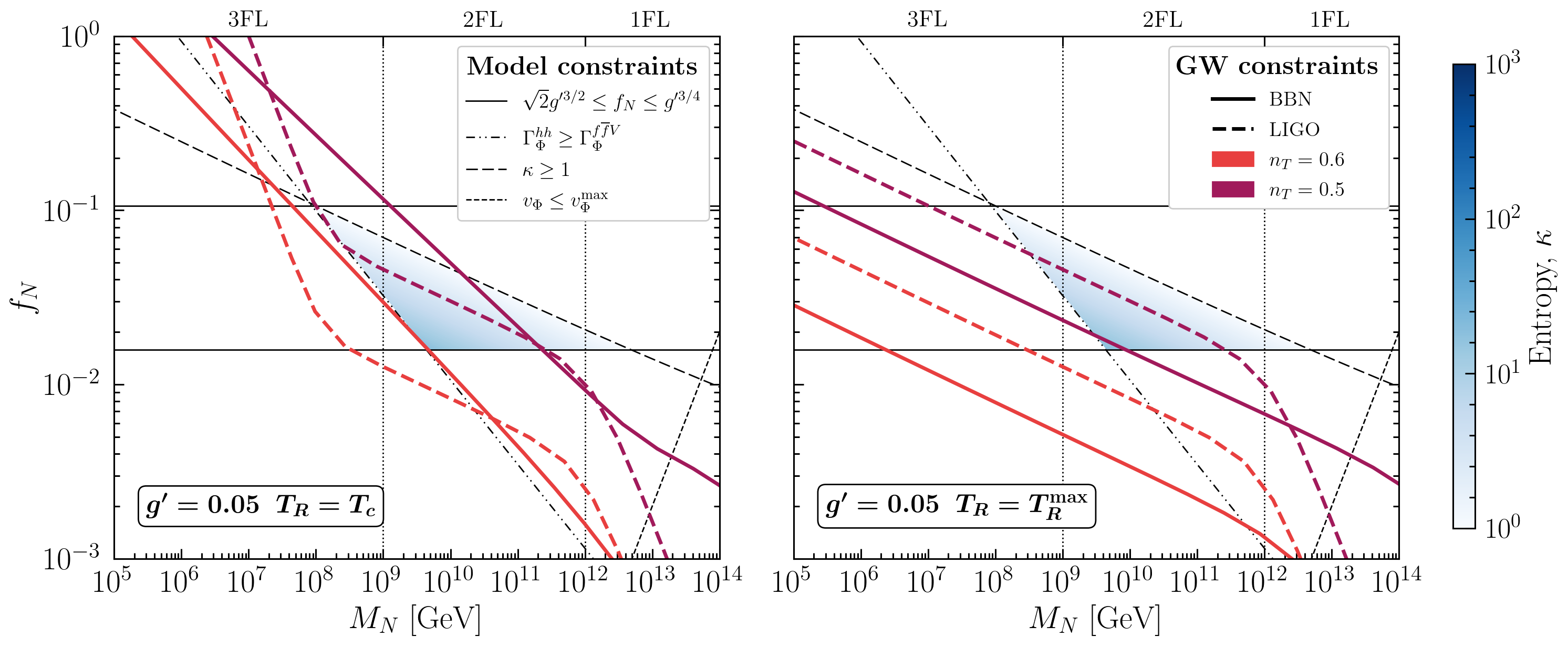}
\caption{Allowed parameter space with $g^\prime=0.05$ for  $T_R = T_c$ (left panel) and $T_R = T_R^{\rm max}$ (right panel). The texture and the color codes for all the curves representing the constraints are the same as in Fig.\ref{fig:GWconstraints}. Note that, compared to the $g^\prime=10^{-2}$ case, smaller values of spectral indices are required because of less entropy production.}
\label{fig:GWconstraints1} 
\end{figure}
\begin{figure}[h!]
\includegraphics[width=\textwidth]{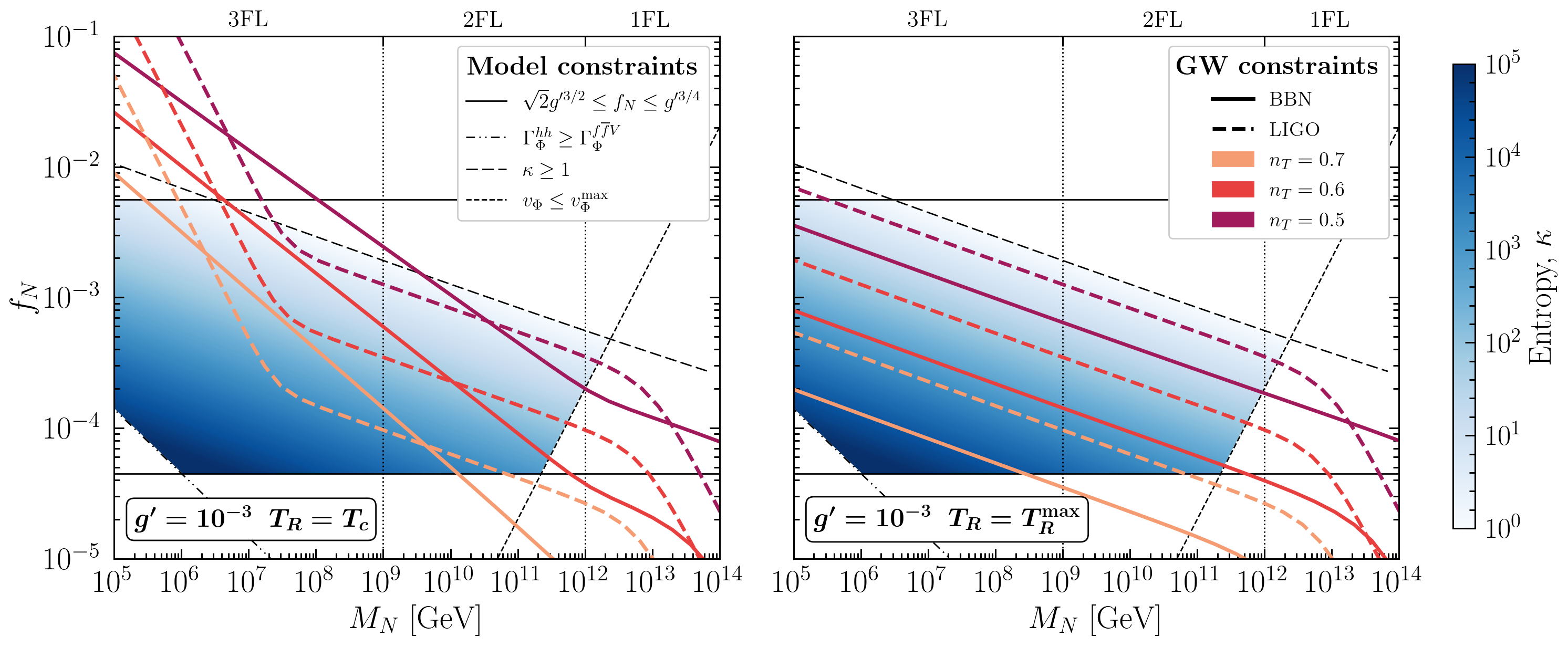}
\caption{Allowed parameter space with $g^\prime=10^{-3}$ for  $T_R = T_c$ (left panel) and $T_R = T_R^{\rm max}$ (right panel). The texture and the color codes for all the curves representing the constraints are the same as in Fig.\ref{fig:GWconstraints}. Note that, compared to the $g^\prime=10^{-2}$ case, larger values of spectral indices are allowed because of larger entropy production.}
\label{fig:GWconstraints3} 
\end{figure}

\bibliography{bibliography}
\end{document}